\documentclass[twocolumn]{aastex62}
\usepackage{rotating}
\usepackage{amsmath} %added on the 14/02/21 for aligned environnement 
\usepackage{amssymb}
\usepackage{xcolor}
\usepackage{hyperref}%added 07/2021 for hyperlinks 
\usepackage{tabularx}
\usepackage{graphicx} %added 12/01/22 for scaling equation 
\usepackage{threeparttable}  %to use tablenotes
\usepackage{xfrac} %added fev 2023 for sfrac
\usepackage{comment}
\usepackage{tablefootnote}

\def\ltsima{$\; \buildrel < \over \sim \;$}
\def\simlt{\lower.5ex\hbox{\ltsima}}
\def\gtsima{$\; \buildrel > \over \sim \;$}
\def\simgt{\lower.5ex\hbox{\gtsima}}
\def\arcsec{\mathop{\rm arcsec}\nolimits} %arc sec
\def\deg{^\circ}
\def\s{\ifmmode \widetilde \else \~\fi}
\def\={\overline}

\def\spose#1{\hbox to 0pt{#1\hss}}

\def\lta{\mathrel{\spose{\lower 3pt\hbox{$\mathchar"218$}}
     \raise 2.0pt\hbox{$\mathchar"13C$}}}
\def\gta{\mathrel{\spose{\lower 3pt\hbox{$\mathchar"218$}}
     \raise 2.0pt\hbox{$\mathchar"13E$}}}
\def\Dt{\spose{\raise 1.5ex\hbox{\hskip3pt$\mathchar"201$}}}    % upper case
\def\dt{\spose{\raise 1.0ex\hbox{\hskip2pt$\mathchar"201$}}}    % lower case

\def\dotsfill{\leaders\hbox to 1em{\hss.\hss}\hfill}
\def\sun{\odot}
\def\earth{\oplus}

\def\FeH{{\rm[Fe/H]}}

\newcommand{\hi}{H\,{\sc i}}
\newcommand{\hii}{H\,{\sc ii}}

\submitjournal{the Astrophysical Journal}

\shorttitle{NGC 3109 satellites}
\shortauthors{Doliva-Dolinsky et al.}

\begin{document}

\title{\bf \large The NGC3109 Satellite System: The First Systematic Resolved Search for Dwarf Galaxies Around a SMC-mass Host}

\correspondingauthor{Amandine Doliva-Dolinsky}
\email{a.doliva-dolinsky@surrey.ac.uk}
\author[0000-0001-9775-9029]{Amandine Doliva-Dolinsky}
\affil{Department of Physics and Astronomy, University of Tampa, 401 West Kennedy Boulevard, Tampa, FL 33606, USA}
\affil{Department of Physics and Astronomy, Dartmouth College, Hanover, NH 03755, USA}
\affil{Department of Physics, University of Surrey, Guildford GU2 7XH, UK}

\author[0000-0001-9649-4815]{Bur\c{c}in Mutlu-Pakdil}
\affil{Department of Physics and Astronomy, Dartmouth College, Hanover, NH 03755, USA}

\author[0000-0002-1763-4128]{Denija Crnojevi\'{c}}
\affil{Department of Physics and Astronomy, University of Tampa, 401 West Kennedy Boulevard, Tampa, FL 33606, USA}

\author[0000-0003-3312-909X]{Dhayaa Anbajagane}
\affil{Department of Astronomy and Astrophysics, University of Chicago, Chicago, IL 60637, USA}

\author[0000-0002-3936-9628]{Jeffrey L. Carlin}
\affil{AURA/Rubin Observatory, 950 North Cherry Avenue, Tucson, AZ 85719, USA} 

\author[0000-0003-1162-7346]{Jonah~Medoff}
\affiliation{Department of Astronomy and Astrophysics, University of Chicago, Chicago, IL 60637, USA}
\affiliation{Department of Physics, Harvard University, 17 Oxford St, Cambridge, MA 02138, USA}

\author[0000-0003-4102-380X]{David J. Sand}
\affiliation{Steward Observatory, University of Arizona, 933 North Cherry Avenue, Tucson, AZ 85721-0065, USA}

\author[0000-0002-9599-310X]{Erik Tollerud}
\affil{Space Telescope Science Institute, 3700 San Martin Drive, Baltimore, MD 21218, USA}

\author[0000-0002-5049-4390]{Sungsoon Lim}
\affiliation{Department of Astronomy, Yonsei University, 50 Yonsei-ro, Seodaemun-gu, Seoul 03722, Republic of Korea}
\affil{Department of Physics and Astronomy, University of Tampa, 401 West Kennedy Boulevard, Tampa, FL 33606, USA}

\author[0000-0001-8354-7279]{Paul Bennet}
\affiliation{Space Telescope Science Institute, 3700 San Martin Drive, Baltimore, MD 21218, USA}

\author[0000-0001-8251-933X]{Alex Drlica-Wagner}
\affil{Astronomy \& Astrophysics, University of Chicago, Chicago, IL 60637 USA}
\affil{Fermi National Accelerator Laboratory, Batavia, IL, USA}

\author[0000-0001-8245-779X]{Catherine E. Fielder}
\affiliation{Steward Observatory, University of Arizona, 933 North Cherry Avenue, Tucson, AZ 85721-0065, USA}

\author[0000-0002-8722-9806]{Jonathan R. Hargis}
\affiliation{Space Telescope Science Institute, 3700 San Martin Drive, Baltimore, MD 21218, USA}

\author[0000-0003-4394-7491]{Kai Herron}
\affil{Department of Physics and Astronomy, Dartmouth College, Hanover, NH 03755, USA}

\author[0000-0001-5368-3632]{Laura Congreve Hunter}
\affil{Department of Physics and Astronomy, Dartmouth College, Hanover, NH 03755, USA}

\author[0000-0002-5434-4904]{Michael G. Jones}
\affiliation{Steward Observatory, University of Arizona, 933 North Cherry Avenue, Tucson, AZ 85721-0065, USA}

\author[0000-0001-8855-3635]{Ananthan Karunakaran}
\affiliation{Department of Astronomy \& Astrophysics, University of Toronto, Toronto, ON M5S 3H4, Canada}
\affiliation{Dunlap Institute for Astronomy and Astrophysics, University of Toronto, Toronto ON, M5S 3H4, Canada}

\author[0000-0002-8040-6785]{Annika H. G. Peter}
\affiliation{CCAPP, Department of Physics, and Department of Astronomy, The Ohio State University, Columbus, OH 43210, USA}

\author[0000-0003-2473-0369]{Aaron J. Romanowsky}
\affil{University of California Observatories, 1156 High Street, Santa Cruz, CA 95064, USA}
\affil{Department of Physics \& Astronomy, San Jos\'e State University, One Washington Square, San Jose, CA 95192, USA}

\author[0000-0002-0956-7949]{Kristine Spekkens}
\affiliation{Department of Physics, Engineering Physics and Astronomy, Queen’s University, Kingston, ON K7L 3N6, Canada}

\author[0000-0002-1468-9668]{Jay Strader}
\affiliation{Center for Data Intensive and Time Domain Astronomy, Department of Physics and Astronomy,\\ Michigan State University, East Lansing, MI 48824, USA}

\author[0000-0003-2892-9906]{Beth Willman}
\affiliation{LSST Discovery Alliance, 933 North Cherry Avenue, Tucson, AZ 85719, USA}

\author[0000-0002-3690-105X]{Julio~A.~Carballo-Bello}
\affiliation{Instituto de Alta Investigaci\'on, Universidad de Tarapac\'a, Casilla 7D, Arica, Chile}

\author[0000-0003-1697-7062]{William Cerny}
\affil{Department of Astronomy, Yale University, New Haven, CT 06520, USA}

\author[0000-0001-5143-1255]{Astha Chaturvedi}
\affiliation{Department of Physics, University of Surrey, Guildford GU2 7XH, UK}

\author[0000-0002-3204-1742]{Nitya~Kallivayalil}
\affiliation{Department of Astronomy, University of Virginia, Charlottesville, VA 22904, USA}

\author[0000-0002-9144-7726]{Clara~E.~Mart\'inez-V\'azquez}
\affiliation{International Gemini Observatory/NSF NOIRLab, 670 N. A’ohoku Place, Hilo, Hawai’i, 96720, USA}

\author[0000-0003-0105-9576]{Gustavo~E.~Medina}
\affiliation{Department of Astronomy and Astrophysics, University of Toronto, 50 St. George Street, Toronto ON, M5S 3H4, Canada}

\author[0000-0002-8282-469X]{Noelia~E.~D.~No\"el}
\affiliation{Department of Physics, University of Surrey, Guildford GU2 7XH, UK}

\author[0000-0002-6021-8760]{Andrew B. Pace}
\affil{Department of Astronomy, University of Virginia, 530 McCormick Road, Charlottesville, VA 22904, USA}

\author[0000-0001-5805-5766]{Alex~H.~Riley}
\affiliation{Institute for Computational Cosmology, Department of Physics, Durham University, South Road, Durham DH1 3LE, UK}

\author[0000-0002-1594-1466]{Joanna D. Sakowska}
\affil{Department of Physics, University of Surrey, Guildford GU2 7XH, UK}

\author[0000-0003-2599-7524]{Adam Smercina}
\affil{Space Telescope Science Institute, 3700 San Martin Drive, Baltimore, MD 21218, USA}

\author[0000-0003-4341-6172]{Kathy~Vivas}
\affiliation{Cerro Tololo Inter-American Observatory/NSF NOIRLab, Casilla 603, La Serena, Chile}

\author[0000-0002-6904-359X]{Monika~Adam\'ow}
\affiliation{Center for Astrophysical Surveys, National Center for Supercomputing Applications, 1205 West Clark St., Urbana, IL 61801, USA}

\author[0000-0003-4383-2969]{Clecio~R.~Bom}
\affiliation{Centro Brasileiro de Pesquisas F\'isicas, Rua Dr. Xavier Sigaud 150, 22290-180 Rio de Janeiro, RJ, Brazil}

\author[0000-0003-1680-1884]{Yumi Choi}
\affil{NSF NOIRLab, 950 N. Cherry Ave., Tucson, AZ 85719, USA}

\author[0000-0001-6957-1627]{Peter S. Ferguson}
\affiliation{DIRAC Institute, Department of Astronomy, University of Washington, 3910 15th Ave NE, Seattle, WA, 98195, USA}

\author[0000-0001-9438-5228]{Mahdieh~Navabi}
\affiliation{Department of Physics, University of Surrey, Guildford GU2 7XH, UK}

\author[0000-0001-6455-9135]{Alfredo Zenteno}
\affiliation{Cerro Tololo Inter-American Observatory/NSF NOIRLab, Casilla 603, La Serena, Chile}

\collaboration{(DELVE Collaboration)}

%\author{friends}

\begin{abstract} 
We report the results of the deepest search to date for dwarf galaxies around NGC3109, a barred spiral galaxy with a mass similar to that of the Small Magellanic Cloud (SMC), using a semi-automated search method. Using the Dark Energy Camera (DECam), we survey a region covering a projected distance of $\sim$70 kpc of NGC 3109 ($D$ = 1.3 Mpc, $R_\mathrm{vir}\sim$ 90 kpc, $M\sim10^8M_\ast$) as part of the MADCASH and DELVE-DEEP programs. Through our resolved and newly designed semi-resolved searches, we successfully recover the known satellites Antlia and Antlia B. We identified a promising candidate, which was later confirmed to be a background dwarf through deep follow-up observations. Our detection limits are well defined, with the sample $\sim 80\%$ complete down to $M_V\sim-$8.0 , and includes detections of dwarf galaxies as faint as $M_V\sim-$6.0. This is the first comprehensive study of a satellite system through resolved star around an SMC mass host. Our results show that NGC 3109 has more bright ($M_V\sim-$9.0) satellites than the mean predictions from cold dark matter (CDM) models, but well within the host-to-host scatter. A larger sample of LMC/SMC-mass hosts is needed to test whether or not the observations are consistent with current model expectations.
\end{abstract}


\section{Introduction} 

\quad In CDM models, dwarf galaxies form within halos of dark matter. While CDM predicts hundreds of these halos to have been accreted and orbiting a central halo \citep{Klypin1999,Moore1999}, not all have formed stars \citep{Bullock2017,Simon2019}. The formation of satellites dwarf galaxies, especially the fainter ones, depends not only on cosmology \citep{Spergel2000,Bode2001} but also on galaxy formation and evolution processes like stellar feedback and reionization \citep{Bullock2000, Benson2002, Somerville2002, Wheeler2015}. Therefore, satellite galaxy observations are essential for testing predictions from cosmological simulations with baryonic physics.

\quad Large photometric surveys \citep[e.g., SDSS, Pan-STARRS1, UNIONS, DES, PAndAS, DELVE;][]{SDSS2003,Pan-STARRS12016, Ibata2017, HSC2018, DES2018, McConnachie2018, Drlica2021} have uncovered many faint satellites of the Milky Way (MW) and M31 through resolved star searches \citep{Martin2013, Laevens2015, Koposov2015, Drlica2015, Drlica2020, Doliva2025}. Along with well-characterized detection limits, the satellites' number, mass, size, and spatial distributions \citep{Willman2002, Drlica2020, Doliva2022, Doliva2023}, make it possible to test cosmological models of structure formation \citep{Koposov2009, Tollerud2008, Kim2018, Nadler2021}. More distant satellite systems of MW-mass hosts in the Local Volume have been studied using resolved, spectroscopic, and unresolved searches \citep{Chiboucas2013,Geha2017, Smercina2018, Bennet2019, Crnojevic2019, Muller2019,Carlsten2021, Mao2024, Mutlu-Pakdil2024}, allowing us to explore satellite properties (e.g., luminosity, size, star formation history) in relation to their host’s environment and merger history. However, the influence of lower mass hosts on their satellite population remains unexplored, as a majority of studies have been focused on MW mass hosts.

\quad The CDM model predicts that even dwarf galaxies should have satellite systems \citep{Munshi2019,Deason2015,Dooley2017b,Santos-Santos2022}. These systems provide a middle ground between the extreme environments of massive galaxy satellites and dwarf galaxies in the field. Indeed, studying isolated, lower-mass hosts is key to understanding how weaker tidal forces and ram pressure stripping affect satellite galaxies, such as their ability to retain gas and continue to form stars \citep{Spekkens2014,Jahn2019, Jahn2022,Garling2024, Jones2024_2}. Recent efforts have also been made to model \citep{Deason2014,Cooper2025} and detect \citep{Jensen2024,Conroy2024, Fielder2025} accreted stellar haloes, offering insight into hierarchical assembly at this mass scale. The Magellanic Clouds (MCs) and M33 are promising candidates for this type of study. Gaia proper motions indicate that $\sim$6 ultra-faint dwarfs are highly likely satellites of the LMC \citep{Kallivayalil2018,Patel2018,Battaglia2022}. However, the fact that the LMC and M33 are within the Local Group makes it difficult to asses the role of the MW’s or M31’s gravitation influence. While a few individual detections of satellites around dwarf galaxies have been reported \citep{Sand2015, Carlin2016, Carlin2021, Carlin2024, Davis2021, Davis2024, Garling2021, McNanna2023, Sand2024}, comprehensive population studies around isolated hosts with masses comparable to the Magellanic Clouds are still lacking. Ongoing efforts aim to address this gap through unresolved searches \citep{Hunter2025, Li2025} and resolved searches, notably within the Magellanic Analog Dwarf Companions and Stellar Halos (MADCASH) survey \citep{Carlin2016, Carlin2024}.

\quad The MADCASH survey and the DEEP component of the DECam Local Volume Exploration Survey \citep[DELVE;][]{Drlica2021} are ongoing sister programs aimed at probing the stellar halos of sub-Milky Way mass hosts in the Local Volume (D$\sim$1-4 Mpc) through deep resolved star observations. These efforts have revealed not only dwarf satellites \citep{Carlin2016, Carlin2021, Carlin2024}, but also stellar substructures and streams around an LMC-mass host \citep{Fielder2025}. In this study, we present the first quantitative luminosity function for dwarf galaxies around an SMC-mass host, NGC 3109 (D = 1.3 Mpc, $M\sim10^8M_\ast$). Sections~\ref{descriptionsurvey} and~\ref{observation} provide an overview of the MADCASH and DELVE-DEEP surveys, as well as the NGC 3109 observations. Section~\ref{searchalgo} details the dwarf galaxy search, and Section~\ref{completeness} explains how we determined the survey completeness. Finally, in Section~\ref{discussion}, we present the resulting luminosity function and discuss the broader properties of NGC 3109’s satellite system. 

\section{Description of Survey} \label{descriptionsurvey}

\begin{deluxetable*}{lcccccc}
\tablecolumns{7}
\tablewidth{0pt}
\tablecaption{List of MADCASH+DELVE-DEEP targets. Stellar masses are taken both from \citet{Dooley2017b} (D) and \citet{Karachentsev2013} (K). In the latter case, the stellar masses are derived from K-band magnitudes using a mass-to-light ratio of 1. The consequent range of possible stellar mass for NGC 3109 and its impact on this study is discussed in Section \ref{sectionLF}. Distances and virial radii are sourced from \citet{McConnachie2012, Karachentsev2013, Mutlu-Pakdil2021}. Predicted numbers of satellites with $M_{\ast} > 10^5~M_\odot$ are estimated in \cite{Dooley2017b}.}
\label{tab:targetlist}
\tablehead{\colhead{Galaxy} & \colhead{$M_{\rm stars} (M_\odot)$ (D)  } & \colhead{$M_{\rm stars} (M_\odot)$ (K)} & \colhead{$D$ (Mpc)} & \colhead{Telescope} & $R_{\mathrm{vir}}$(kpc) & \colhead{$N_{\rm sat,exp}$} }
\startdata
Sextans B  & $4.4\times10^7$ & $5.9\times10^7$ & 1.4& DECam& $\sim90$ & 0-2\\
Sextans A  & $5.2\times10^7$& $3.0\times10^7$ & 1.4& DECam & $\sim90$ &0-2\\
NGC 3109  & $7.6\times10^7$& $3.5\times10^8$ & 1.3& DECam & $\sim90$ & 1-3  \\
IC 4662  & $1.9\times10^8$ & $4.7\times10^8$ & 2.4& DECam & $\sim90$ &1-4\\
IC 5152  & $2.7\times10^8$& $5.1\times10^8$ & 2.0& DECam & $\sim90$  &1-4\\
***SMC  & $7.0\times10^8$ & $6.3\times10^8$ & 0.06& -- & -- & 1-5 \\
NGC 4214  & $1.0\times10^9$ & $9.7\times10^8$ & 3.0& Subaru& $\sim100$ &  1-5 \\
NGC 300  & $2.6\times10^9$ & $2.6\times10^9$ & 2.1& DECam& $\sim120$ & 2-7 \\
***LMC  & $2.6\times10^9$ &  $2.5\times10^9$ & 0.05& -- & -- &  2-7 \\
NGC 55 &  $3.0\times10^9$ & $2.9\times10^9$ & 2.1& DECam & $\sim120$ &  2-8 \\
NGC 247 & $3.2\times10^9$ & $2.8\times10^9$  & 3.7& Subaru& $\sim120$ &  3-8 \\
NGC 4244 & $3.5\times10^9$ & $3.3\times10^9$ &4.3 & Subaru & $\sim120$ & 2-6\\
NGC 2403 & $7.2\times10^9$  & $6.9\times10^9$  & 3.2& Subaru& $\sim140$ &  4-11 \\
\enddata
\end{deluxetable*}

\begin{figure*}[t]
    \centering
    \includegraphics[width=0.8\textwidth]{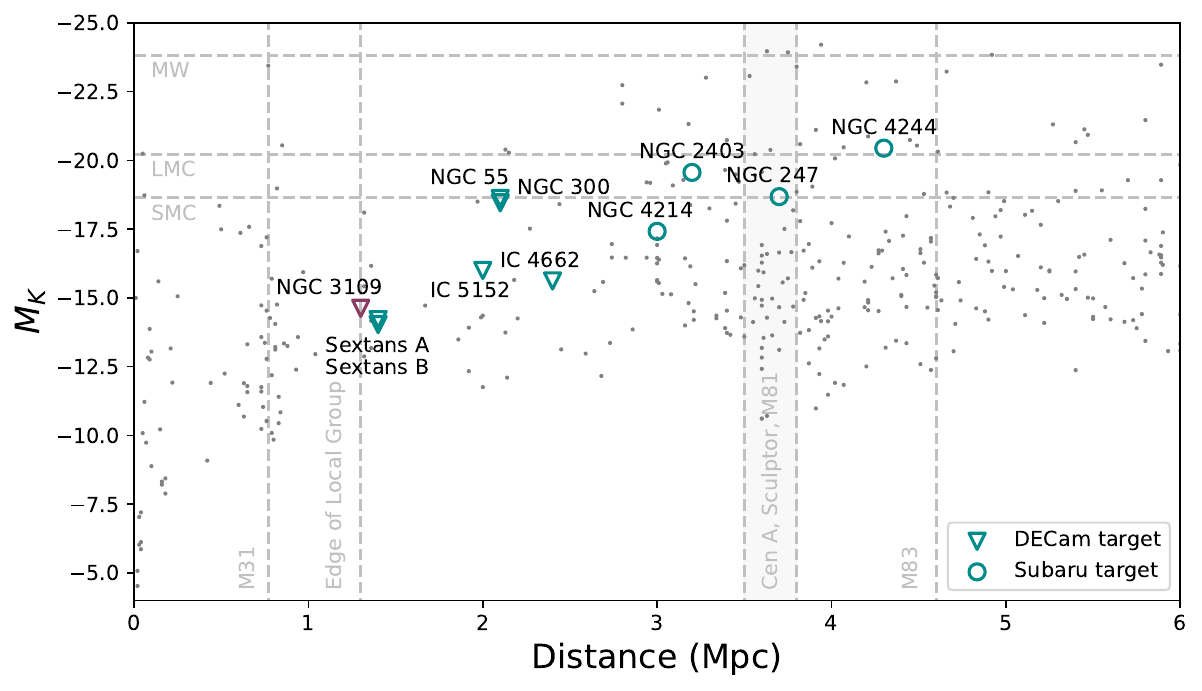}
    \caption{Absolute K-band magnitude as a function of distance for nearby known galaxies \citep[gray dots;][]{Karachentsev2013}. For reference, the absolute K-band magnitude of the SMC, LMC, and MW are highlighted (grey dashed lines). Our target hosts from Tab.~\ref{tab:targetlist} are represented by circle and triangle symbols (see legend), with the host of interest in this paper highlighted in burgundy. These hosts span a stellar mass range from approximately 1/20 $M_{\textrm{SMC}}$ to ~3 $M_{\textrm{LMC}}$. The hosts are within 4.5 Mpc from the MW. Although other galaxies meet the stellar mass and distance criteria, some were excluded due to being too far north for observation or impacted by high extinction.}
    \label{fig:massedistance}
\end{figure*}

\begin{figure}[t!]
    \centering
    \includegraphics[width=0.48\textwidth]{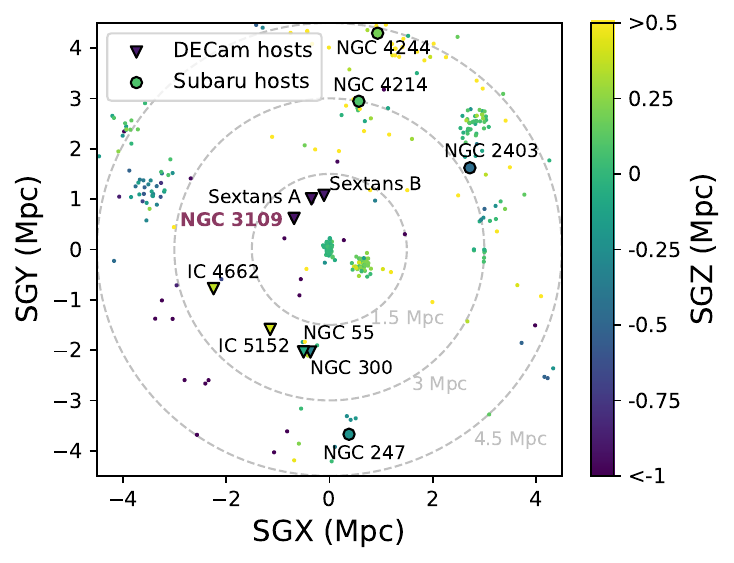}
    \caption{Spatial distribution of nearby dwarf galaxies projected in the Cartesian supergalactic SGX-SGY plane, color-coded by their SGZ value; data from the Local Volume Database \citep{Pace2024}. Circle and triangle symbols highlight our target hosts, with the host of interest in this paper highlighted in burgundy; note that the hosts appear to be relatively isolated.}
    \label{fig:targetlist}
\end{figure}

\quad NGC~3109 observations were obtained through the MADCASH survey, designed to address the observational gap in the study of faint satellite dwarf galaxies around SMC-LMC mass dwarf galaxies. The goal is to derive the global and individual properties of such satellite systems (e.g., number of dwarfs, luminosity function, spatial distribution, size-luminosity relation, star formation properties) and explore the implications for galaxy formation and evolution in the $\Lambda{\rm CDM}$ framework \citep{Dooley2017a,Dooley2017b,Santos-Santos2022}. The survey obtained deep, homogeneous, wide-field photometric data of regions surrounding Local Volume galaxies with masses similar to those of the LMC and SMC. As targets are located in both hemispheres, both the Dark Energy Camera \citep[DECam;][]{Flaugher2012} on the Blanco 4-meter Telescope and Hyper Suprime-Cam \citep[HSC;][]{Miyazaki2012} on the 8.2-m Subaru telescope, which have respective fields of view of 3 and 1.8 deg, were used for the MADCASH observations. Later, MADCASH partnered with DELVE-DEEP to expand the sample of targets using DECam, employing a similar observing strategy to the original approach.

\quad In a combined effort of the MADCASH and DELVE-DEEP surveys, we target the halos of 11 Local Volume dwarf galaxies with stellar masses ranging from $4.4\times10^7$ ($\sim$1/20$M_{\textrm{SMC}}$) to $7.2\times 10^{9} M_{\odot}$ ($\sim$3$M_{\textrm{LMC}}$) and located within 4~Mpc, aiming to resolve their stellar populations. Observations are conducted in the $g$ and $r$ bands or $g$ and $i$ bands, reaching a depth of $\sim$1.5-2 magnitudes fainter than the tip of the red giant branch (TRGB), enabling the detection and characterization of dwarf galaxies with luminosities at least as faint as $M_V \sim -7$. A list of the eleven targeted hosts is given in Table~\ref{tab:targetlist}. Figure~\ref{fig:massedistance} shows the K-band magnitude distribution of the hosts as a function of distance. The eleven selected target galaxies have estimated virial radii of about 90-140 kpc \citep{Guo2013, Mutlu-Pakdil2021}. This allows their halos to be effectively imaged with about 10 telescope pointings each, using the Blanco/DECam or Subaru/HSC wide-field imagers. These targets are at distances where we can resolve individual red giant branch (RGB) stars, and while additional galaxies meet the mass and distance criteria, some have been excluded because they are too far north to be observed or affected by high extinction. 

\quad Figure~\ref{fig:targetlist} shows the 3D spatial distribution of the hosts in supergalactic coordinates, showing that they are relatively isolated; in other words, they are not satellites of a higher-mass host. This provides ideal conditions for studying satellite systems that are not influenced by a larger host. NGC~3109 is the most massive galaxy in the Antlia-Sextans dwarf galaxy association, which also includes Sextans A, Sextans B, Antlia, Antlia B, and Leo P. While these galaxies share a common association, only Antlia and Antlia B are classified as satellites as they fall within the virial radius of NGC~3109. We note that Antlia B was identified through the MADCASH dataset \citep{Sand2015}. Table~\ref{ta:tableprop} presents NGC~3109 and its known satellite galaxies' properties.

\section{Observations and data reduction} \label{observation}

\begin{deluxetable*}{lccccccc} 
\tablecolumns{8}
\tablewidth{0pt}
\tablecaption{Properties of known dwarfs in the NGC 3109 satellite system extracted from \cite{McConnachie2012,Sand2015,Hargis2019}. Stellar masses are derived from $M_V$ assuming a mass-to-light ratio of 1. \label{ta:tableprop}}
\tablehead{\colhead{Galaxy} & \colhead{R.A. (J2000)} & \colhead{Dec (J2000)} & \colhead{$D$ (kpc)} & D$_{\textrm{NGC3109}}$(deg) & \colhead{$M_V$} & \colhead{$M_\ast$} ($M_\odot)$  & \colhead{r$_\textrm{h}$(pc)}  } 
\startdata
NGC~3109 & 10h03m06.9s & -26d09m35 & 1300$\pm$48 & &-14.9$\pm$ 0.1  & $\sim10^8$  &1626$\pm$71 \\
Antlia & 10h04m04.1s & -27d19m52s &1349$\pm$62 & 1.2 &-10.4$\pm$0.2 &  $\sim10^6$&471$\pm$52\\
Antlia~B & 09h48m56.1s & -25d59m24s & 1350$\pm$60 & 3.2  &-9.7$\pm$0.6 & $\sim10^{5.8}$&273$\pm$29 \\
\enddata
\end{deluxetable*}

\begin{figure*}[t]
\centering
\includegraphics[width=\textwidth]{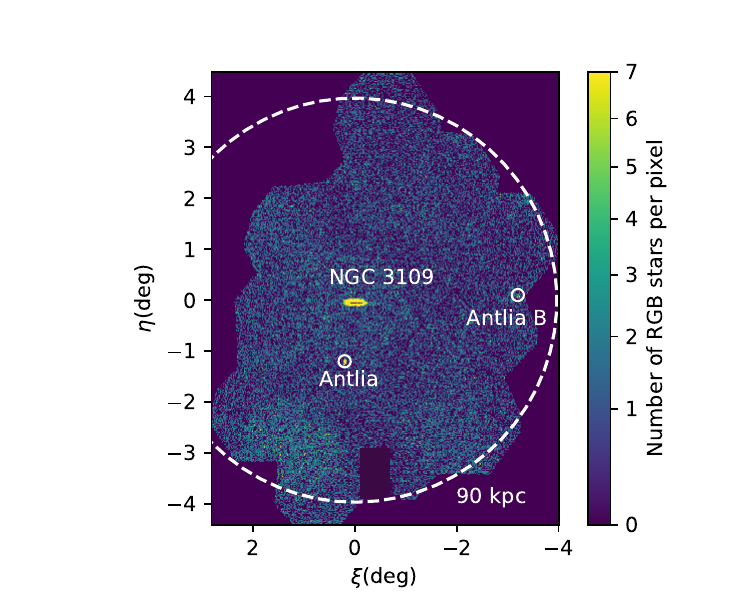}
\caption{RGB map of the NGC~3109 halo from the MADCASH$+$DELVE-DEEP surveys. The RGB filtering is described in Figure~\ref{cmdMADCASH}. The footprint is divided into pixels of $\sim$1.2\arcmin $\times$ 1.2\arcmin.  We overlay the estimated virial radius of NGC 3109 \citep[90 kpc;][]{Mutlu-Pakdil2021}. The dark rectangle to the south marks a region with unreliable data. The footprint is fairly uniform in depth, though NGC 3109 and two southern regions receive extra coverage and depth from DELVE. The two known satellites appear as clear spatial overdensities of old metal-poor stars.}
\label{rgbmap}
\end{figure*}

\begin{figure*}[t]
\centering
\includegraphics[width=0.7
\textwidth]{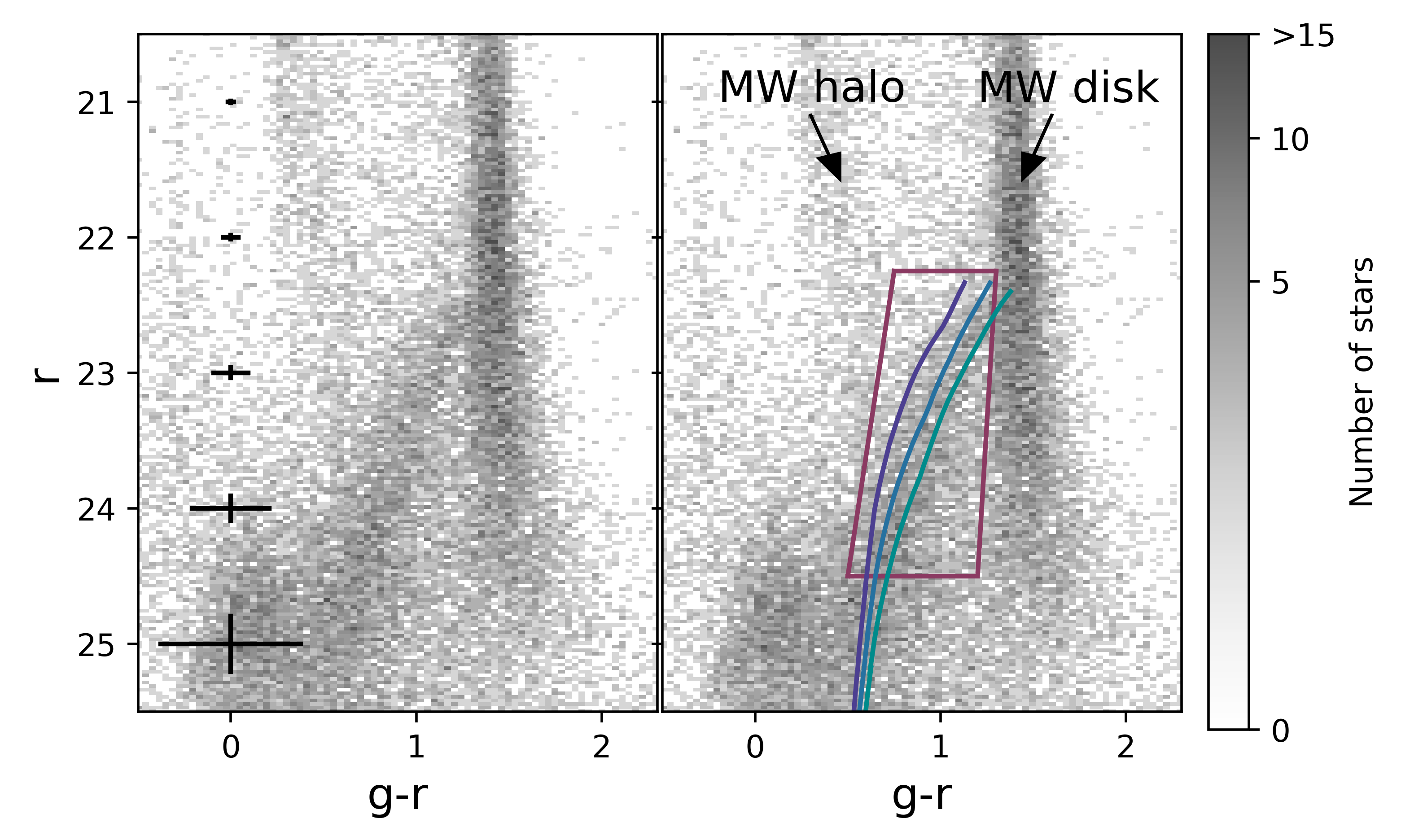}
\caption{Both panel shows the same color magnitude diagram (CMD) for a region of $\sim$ 1 square degree close to NGC~3109. The CMD is binned into 0.03 mag and 0.04 mag pixels respectively in magnitude and color and displayed on a power law scale with an exponent of 0.4. On the left panel, we display the typical photometric uncertainties. On the right panel, Padova isochrones \citep{Bressan2012}, shifted to NGC~3109's distance, are superimposed (blue lines; age=10 Gyr and \FeH$=-2.3, -1.7, -1.4$ from left to right). The burgundy box represents our selection of RGB stars, notably used to generate the RGB map for the NGC 3109 survey shown in Figure \ref{rgbmap}. The foreground contamination from MW stars is high, with the disk component slightly overlapping  with the RGB population of NGC 3109. We therefore design the selection box to minimize contamination.}
\label{cmdMADCASH}
\end{figure*}

\quad The halo of NGC 3109 was observed with the DECam instrument (2015A-0130; PI: D. Crnojević). The footprint of the survey is displayed in Figure~\ref{rgbmap}. The survey's 15 pointings probed a region within a projected radius of $\sim$50 kpc north-east of the host and of $\sim$90 kpc elsewhere. Data were obtained with $7\times300$~s and $7\times150$~s exposures in the $g$ and $r$ bands with good seeing (FWHM $< 1.0$ arcsec). Between each exposure, small dithers were used to cover the gaps between CCDs.

\quad The photometric catalog is derived from DELVE, which reprocesses our MADCASH dataset together with more recent and archival DECam observations (adding 3$\times$300~s exposures in $g$ and $r$ for the whole field of view).  The photometry is obtained following the DES Year 6 (Y6) pipeline through multi-band and multi-epoch fitting \citep{Hartley2022}. This produces a Single-Object Fit (SOF) catalog with more precise photometry and better star-galaxy separation than the coadded catalog, thanks to improved PSF modeling. We separate stars from background galaxies using a morphological classification designed for DES Y6. A more detailed description of both processes is given in \cite{Drlica2022,Tan2024,Bechtol2025,Anbajagane2025}. The resulting catalog of stars is then corrected for extinction using the $E(B-V)$ maps from \cite{Schlegel1998} (with a typical $E(B-V)$ of 0.05 mag in this region) and the coefficients derived for DES in \cite{Schlafly2011}. The survey median $10\sigma$ depth is 24.9, which corresponds to the magnitude where the signal-to-noise ratio reaches 10 for a point source.

\quad The color-magnitude diagram (CMD) of stars within 1 square degree of NGC~3109 is presented in Figure~\ref{cmdMADCASH}. Overlayed are the Padova isochrones in the DECam filter system \citep{Bressan2012} for \FeH$= -2.3, -1.7, -1.4$ (from left to right) with a fixed age of 10 Gyr to represent the position of the RGB stars at the distance of NGC 3109. The regions corresponding to stellar contamination from the MW disk and the MW halo are highlighted in this CMD. One can notice that the contamination by MW disk stars overlaps slightly with the TRGB stars at the distance of NGC~3109. Since it can artificially increase the number of detected RGB stars, MW contamination results in a higher rate of false detections when using matched filter detection methods. Therefore, we adjust the detection threshold to minimize false detections caused by contamination (see further details in Section 4).

\quad Figure~\ref{rgbmap} presents the RGB map of our NGC~3109 survey. The selection box used for RGB filtering, shown in Figure~\ref{cmdMADCASH}, is designed to minimize the contamination of stars in the MW disk. Antlia \citep{Whiting1997} and Antlia B \citep{Sand2015}, the two previously known satellites, appear as clear overdensities in this map. The halo is devoid of other obvious stellar substructures. The MADCASH survey depth around NGC 3109 is relatively uniform across the entire footprint, with a pointing on NGC 3109 and two to the south receiving additional coverage and depth. As a result, these areas have cumulative exposure times that are approximately 1.5 to 2 times longer, resulting in deeper observing field coverage.

\section{Dwarf Satellite Search} \label{searchalgo}

\quad At the distance of NGC 3109 (1.3 Mpc), bright-compact dwarf galaxies have crowded central regions which hinder our photometric pipeline’s ability to distinguish individual stars. As a result, SourceExtractor \citep{Bertin1996} misidentifies those crowded regions as multiple separate low surface brightness sources, effectively ``shredding" the light. As there are fewer resolved stars, the observed spatial overdensity of RGB stars is weaker for such semi-resolved objects. We note that employing a PSF-fitting code such as DAOPHOT could have improved the detection of resolved dwarf galaxies. However, applying such methods across the entire survey would have reduced automation and be significantly more time-consuming. The impact of this ``shredding" effect is discussed in more detail in \cite{Medoff2025}, and motivated the combination of resolved and semi-resolved searches for dwarf galaxies to achieve maximum detection efficiency.

\subsection{Visual Search}
\quad In the early stage of the survey, the MADCASH coadded images were visually inspected for dwarf galaxies, leading to the detection of Antlia B \citep{Sand2015}. Five other candidates were identified in unresolved light, but they later turned out to be background galaxies with our Hubble Space Telescope (HST) follow-up observations (see Appendix~\ref{HST}). This highlights the importance of follow-up observations for unresolved dwarf candidates. 

\begin{figure*}[t]
\centering
\includegraphics[width=\textwidth]{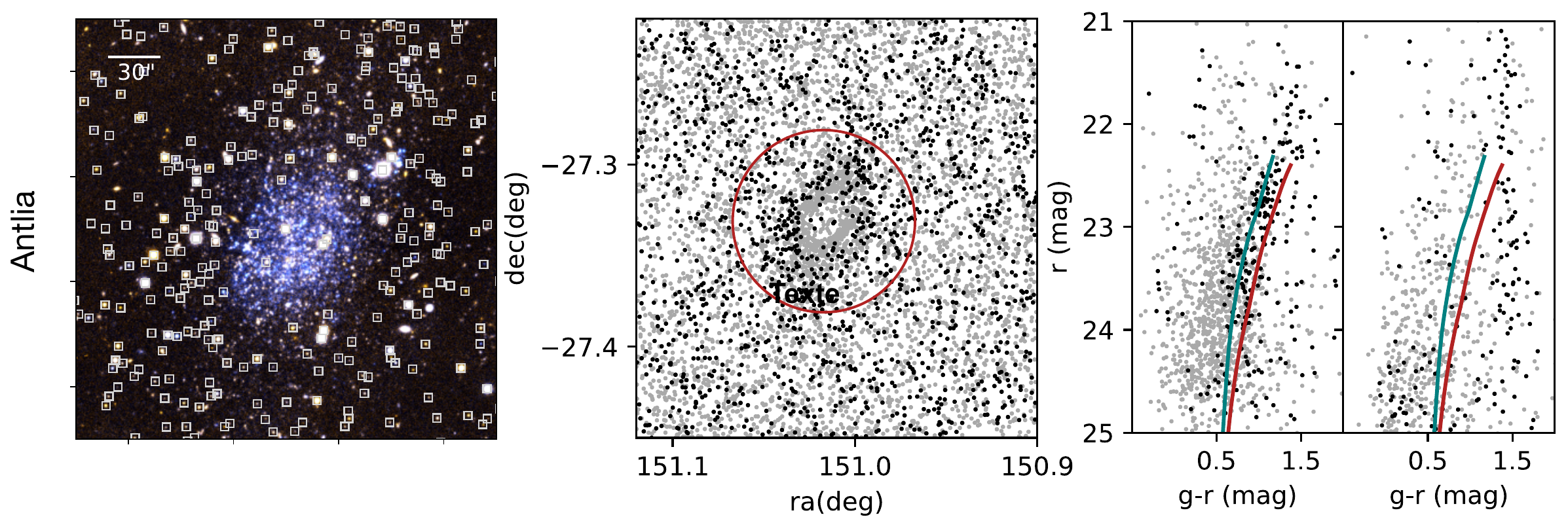}
\includegraphics[width=\textwidth]{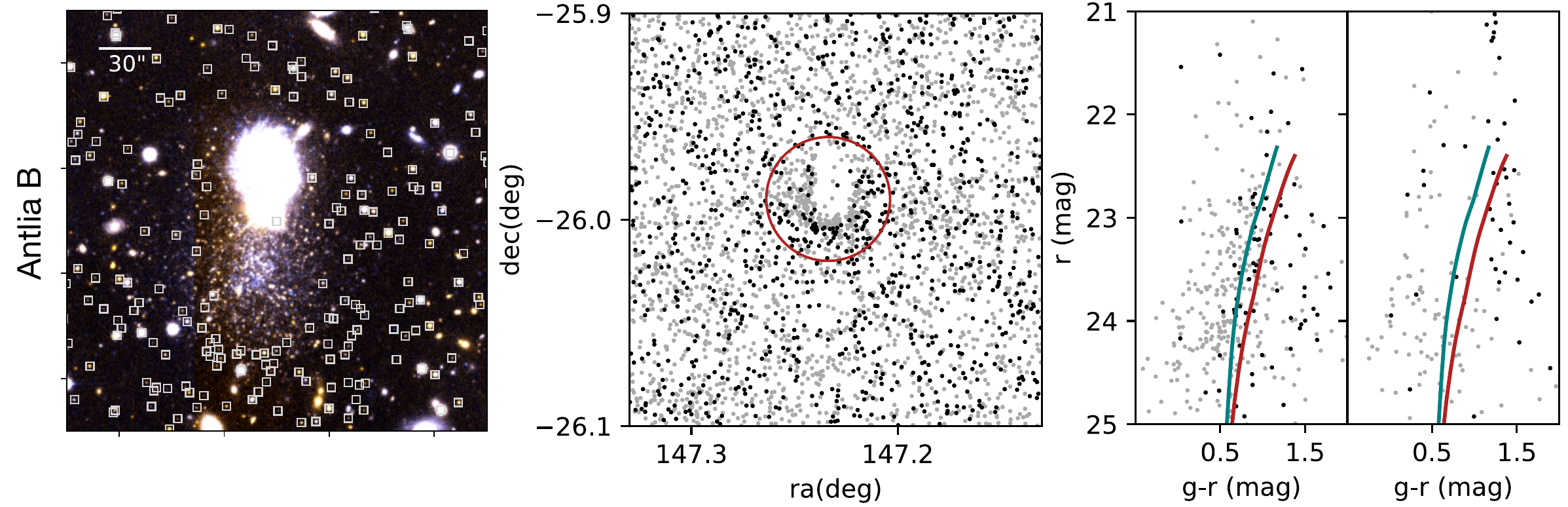}
\caption{\textit{Top, from left to right:} the image, resolved stellar spatial distribution, CMD for the known dwarf galaxy Antlia and CMD for a randomly selected
background region. In the image, RGB stars recovered by the reduction pipeline are marked with white squares. The spatial distribution panel shows the same RGB stars in black on a more zoomed-out scale, while grey dots represent all objects in the catalog. We note that most of the image is contained within the red circle. In the CMD, stars are selected from within the red circle in the spatial distribution. We overlay Padova isochrones \citep[age = 10 Gyr and $\FeH = -2$ to $-1.4$ from left to right;][]{Bressan2012}. \textit{Bottom:} The same panels are presented for Antlia B. For both dwarf galaxies, their crowded center complicates obtaining reliable photometry in their central regions. }
\label{antlia_antliab}
\end{figure*}

\subsection{Resolved Dwarf Search}

\quad In this section, we present the resolved method for detecting NGC 3109 dwarf galaxies. Given the typical composition of these galaxies, we search for spatial overdensities of old \citep[$\sim 10$ Gyr old;][]{Weisz2019} metal-poor \citep[\FeH$\sim-2$;][]{Kirby2013} RGB stars at the distance of NGC 3109 (1.3 Mpc; see Figure \ref{cmdMADCASH}). We use a maximum-likelihood-based match filter method \citep{Kepner1999, Rockosi2002} with the search done identically to \cite{Mutlu-Pakdil2021}, and we summarize this process and our choices in the following paragraph. 

\quad We opt not to apply star-galaxy separation to the catalog for our resolved dwarf search. Figure~\ref{antlia_antliab} shows the images of Antlia and Antlia~B with the resolved RGB stars highlighted with white boxes (left panels), the source's spatial distributions (middle panels), and the CMDs obtained (right panels). All sources are plotted in grey, while we highlight the one selected as stars in black. We note that since NGC~3109 dwarf satellites might appear as crowded systems, obtaining accurate photometry for stars near their centers can be challenging. This difficulty means that our typical morphological star-galaxy separation criteria inadvertently exclude stars that actually belong to these dwarf galaxies, thereby reducing the detection signal. Additionally, our artificial dwarf tests (see Section~\ref{completeness}) demonstrate that the matched filter technique yields better results without excluding stars based on star-galaxy separation. 

\quad The dwarf galaxy component is generated from a well-populated simulated CMD based on an old, metal-poor PARSEC isochrone and luminosity function \citep[10 Gyr, $\FeH=-2$, 1.3 Mpc;][]{Bressan2012}, incorporating estimated completeness and photometric uncertainties (we adopt the values derived for NGC 55 in \citealt{Medoff2025}). Since the positions of the dwarf galaxies are not known, the background CMD is assumed to be the mean stellar density of the field of interest. We choose fields of $1\deg\times1\deg$ so that any potential dwarf signal is negligible for the background calculation as we expect dwarf to have $r_\textrm{h}\leq0.05\deg$. We bin the modeled dwarf and background CMDs into 0.15 mag $\times$ 0.15 mag color–magnitude bins with $-0.5<g-r<2$ and $20<r<24.5$. The observed stellar catalog is spatially binned into 20 arcsec pixels and filtered using the same $g-r$ and $r$ limits as the model CMDs. Using the maximum-likelihood matched filter method \citep{Kepner1999, Rockosi2002}, we generate a spatial significance map of the dwarf signal relative to the background. Finally, we smooth this map with a Gaussian kernel matching the 20 arcsec pixel size.

\quad We use a significance threshold of 5$\sigma$ RGB overdensity over the background to classify outputs of the matched filter as detections. This threshold was chosen to optimize detection rates based on our completeness tests (see Section~\ref{completeness}), while keeping false positives at a manageable level. The resulting 1,725 detections are mostly false positives, due to factors such as MW contamination and background galaxy clusters. We use a visual classification process to identify follow-up worthy candidates among these detections (see Section \ref{searchresults}). Both known dwarf galaxies of NGC 3109, Antlia and Antlia B, are recovered with a high significance ($>20\sigma$). % We note that our CMD for Antlia B appears less populated than previously reported by \citet{Sand2015}, despite using the same imaging data. The primary reason for this difference is the use of distinct photometric pipelines: our study employs a SExtractor-based detection pipeline, whereas the original discovery utilized DAOPHOT, which is better suited for crowded fields but is also more computationally intensive than our method.

\subsection{Semi-resolved search}

Our automated semi-resolved detection technique, inspired by \cite{Tanoglidis2021}, uses SExtractor parameter cuts and has been developed for the search for dwarf galaxies around NGC 55 \citep{Medoff2025}. In the case of crowded dwarfs, SExtractor tends to ''shred" the central crowded region into multiple extended low-surface-brightness sources \citep{Prole2018}. Our method leverages SExtractor parameters \citep{Bertin1996} to filter data and enhance the detection of spatial overdensities of faint extended features, specifically optimizing the identification of the shredded unresolved components within crowded dwarf galaxies. We note that we first use the star/galaxy separation criteria from \cite{Tan2024} to remove all the source identified as stars. Key parameters used in this analysis include the spread from model fitting, \texttt{SPREAD$\_$MODEL}, the error on this parameter \texttt{SPREADERR$\_$MODEL}, the fraction-of-light radius \texttt{FLUX$\_$RADIUS}, the effective modeled object surface brightness \texttt{MU$\_$EFF$\_$MODEL} and the $g-r$ color obtained with SExtractor \texttt{MAG\_AUTO} values. 
The cuts implemented are: 
\begin{itemize}
    \item \texttt{SPREAD$\_$MODEL}+5/3$\times$\texttt{SPREADERR$\_$MODEL}$>0.007$
    \item $5<$\texttt{FLUX$\_$RADIUS}$<20$
    \item $24.2<$\texttt{MU$\_$EFF$\_$MODEL}$<31.2$
    \item $-0.1<g-r<1.4$
\end{itemize}
Objects selected by these cuts are binned into 1\arcmin$\times$1\arcmin ~ to create a density map of low surface brightness features. We then apply the \texttt{find$\_$peaks} function from \texttt{photutils} \citep{Astropy2013, Astropy2018, Astropy2022} on this density map to identify overdensities that may correspond to extended, crowded dwarf galaxies. We use a 3$\sigma$ overdensity as our detection threshold, chosen to optimize completeness while maintaining manageable false positives based on our artificial dwarf injections (see Section~\ref{completeness}). This results in a sample of 1,331 detections, most of which are false positives caused by foreground stars or background galaxies. We note that Antlia and Antlia B are both recovered through this method. 

\subsection{Search results} \label{searchresults}

\begin{figure*}[t]
\centering
\includegraphics[width=\textwidth]{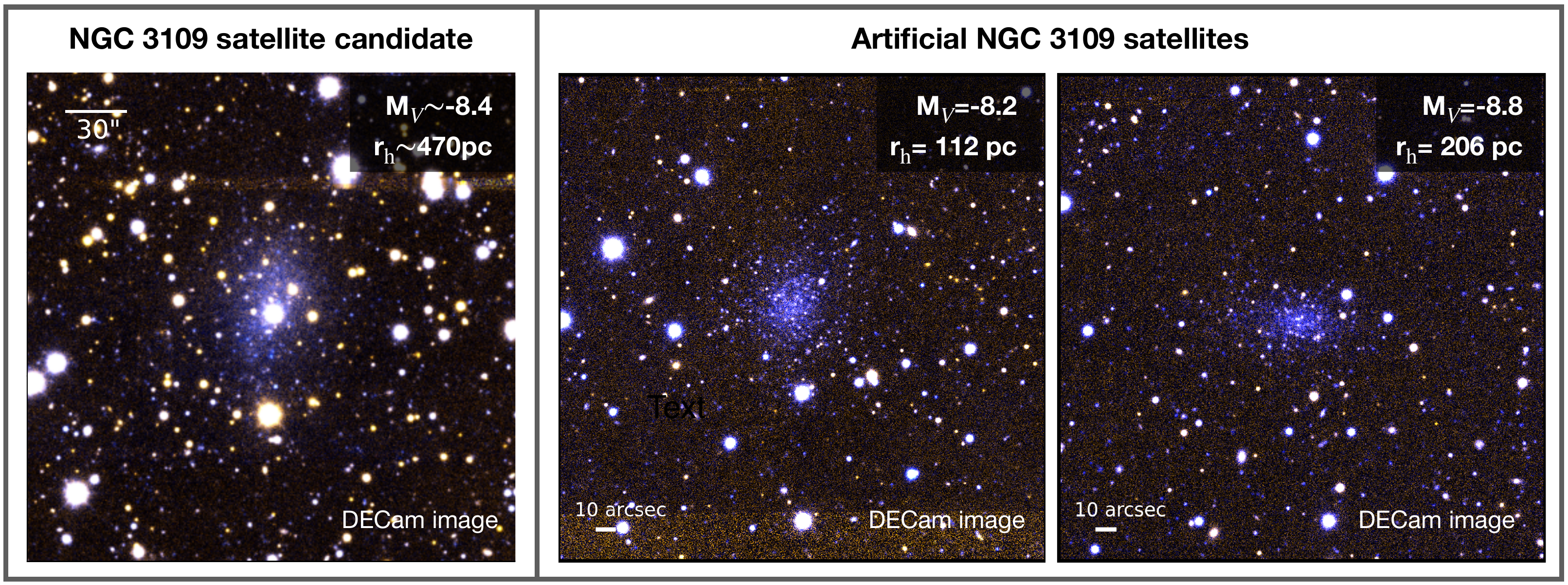}
\caption{DECam $gri$ images of the candidate NGC~3109 dwarf detected by our semi-resolved search algorithm. Assuming a distance similar to NGC 3109, we estimate its V-band magnitude to be $M_V=-8.4$ and its half-light radius to be $r_\textrm{h}=470$ pc. \textit{Right Panel:} DECam $gri$ images of two artificial dwarf galaxies at the distance of NGC~3109 with $M_V=-8.2$ and $M_V=-8.8$ and $r_\textrm{h}=112$ pc and $r_\textrm{h}=206$ pc, respectively (see Section~\ref{completeness}). The candidate and fake dwarfs are patchy, showing both resolved stars and unresolved light. Its lower resolution compared to the fake dwarf suggests that it is a background, isolated dwarf galaxy.}
\label{imagecandidate}
\end{figure*}

\quad After implementing both search methods separately, we identify 1,725 candidates with the resolved method and 1,331 with the semi-resolved method. Upon cross-referencing to eliminate duplicates, we compile a final tally of 2,925 unique candidates.
We then generate diagnostic plots, including the CMD, spatial distribution, radial density profile, image, and stellar luminosity function, as shown in \cite{Carlin2024}. We conduct visual inspections of these plots via a private project on the citizen science platform, Zooniverse\footnote{\url{https://www.zooniverse.org}}, searching for signs that might indicate potential dwarf galaxy candidates. We specifically look for features like RGBs in the CMD, overdensities in the stellar spatial distribution, or unresolved light in the images. Our team evaluates whether each detection qualifies as a candidate for follow-up or as a non-candidate, prioritizing sample purity over completeness. Each detection is reviewed by three team members, and only those that receive unanimous approval for follow-up are considered candidate satellites. Among the 2,925 detections, our team consistently classifies seven as candidate dwarf galaxies. Two of these are Antlia and Antlia B. Three are attributed to MW contamination or cirrus upon further investigation of the CMDs and images. One matches a candidate previously identified visually and already ruled out based on our HST observations (see Appendix~\ref{HST}, \citealt{Hargis2019}). The last detection appears as a possible new satellite dwarf galaxy of NGC~3109. It is located at right ascension 148.683$\deg$ and declination -28.515$\deg$. This corresponds to the concurrent and independent discovery reported in \citet{Karachenstev2024}, where it is named LDD 0954-28. We present its image alongside those of comparable simulated dwarf galaxies (see Section~\ref{completeness}) in Figure \ref{imagecandidate}. Both the candidate and the fake dwarf galaxies appear semi-resolved (i.e, resolved stars on top of unresolved light), however, LDD 0954-28 seems less patchy, indicating it might be more distant which motivated obtaining follow-up imaging (see next Section~\ref{followup}). 

\quad Each candidate was reviewed by three members of the team on the Zooniverse platform. We reviewed all candidates where two out of three reviewers flagged it as being of interest. Among these, two stood out as potential dwarf galaxies, although they appear unresolved and to be background objects (see Appendix \ref{backgrounddetection}).

\subsection{Gemini follow-up imaging of LDD 0954-28} \label{followup}

\begin{figure*}[t]
\centering
\includegraphics[width=\textwidth]{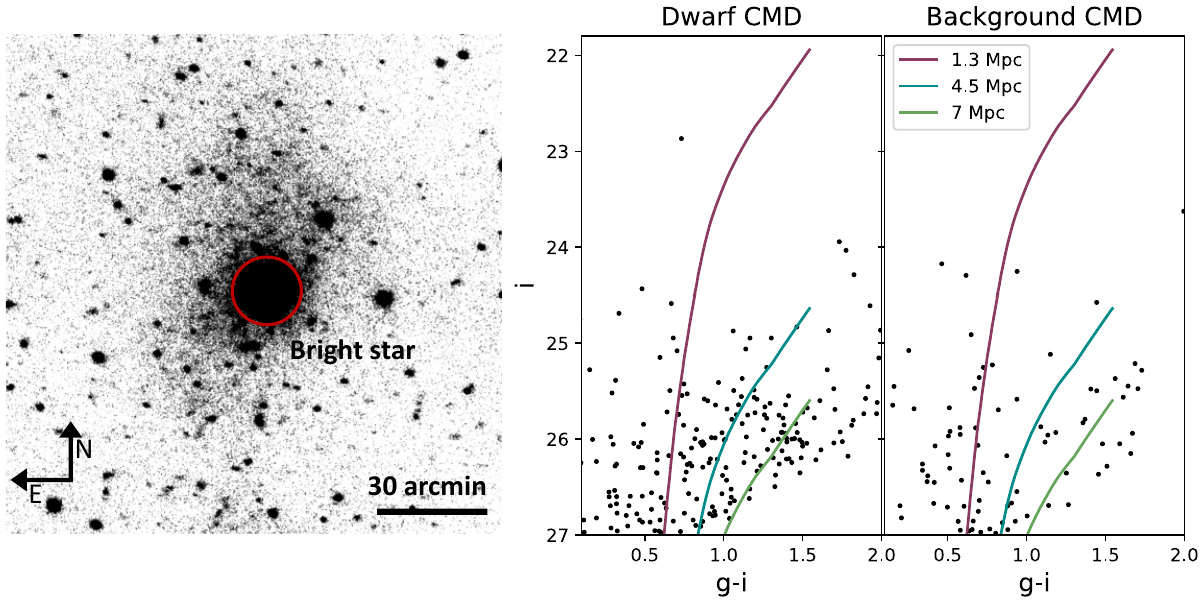}
\caption{\textit{Left panel:} Gemini GMOS g-band image of the candidate NGC~3109 satellite, LDD 0954-28. The DECam gri image of LDD 0954-28 is shown in Figure~\ref{imagecandidate}. We circle in red the bright foreground star projected onto the dwarf. \textit{Right panel:} CMD of LDD 0954-28 and of a randomly selected
background region of the same size. We overlay Padova isochrones (10 Gyr old, $\FeH=-2$ and D=1.3, 4.5 and 7 Mpc). The absence of bright RGB stars along the isochrone at 1.3 Mpc, despite excellent seeing conditions, suggests the system is a background dwarf galaxy.}
\label{geminicand}
\end{figure*}

\quad  We performed deep $g$- and $i$-band follow-up imaging using the Gemini Multi-Object Spectrograph \citep[GMOS;][]{Hook2004} on the 8.1-meter Gemini South telescope as part of program GS-2025A-Q-143 (PI: Doliva-Dolinsky). To mitigate the effect of the bright star at the center of the dwarf, we obtained 16 exposures of 120 seconds each for both the $g$ and $i$ bands, applying small dithers between exposures. The raw data were processed using the DRAGONS pipeline from the Gemini Observatory, following the methodology outlined in \cite{Sand2024}. To extract photometric measurements, we applied point-spread function (PSF) fitting on the combined GMOS images using DAOPHOT and ALLFRAME \citep{Stetson1987}, following the approach described by \cite{Mutlu-Pakdil2018}. The resulting photometry was calibrated against point sources from the DELVE DR3 catalog \citep{Tan2024}. Additionally, we accounted for extinction on an individual star basis following the corrections from \cite{Schlafly2011}.

\quad Figure~\ref{geminicand} shows the $g$-band image of LDD 0954-28 obtained with GMOS. Although the central bright clump could have been a star cluster or \hii\  region, Gaia \citep{Gaia2021} measured its proper motion, confirming it as a foreground star. We also present the CMD for both LDD 0954-28 and the background population. We select stars for the CMD to prevent false detections around the bright foreground star. We overlay Padova isochrones representing an old (10 Gyr), metal-poor ($\FeH=-2$) population at different distances on the CMDs. At the distance of NGC~3109 (1.3 Mpc), the expected bright stars along the isochrone are missing, and there is no overdensity in the fainter region. If considered a satellite of NGC~3109, GALFIT \citep{Peng2002} estimates LDD 0954-28 parameters as $M_V=-8.4$ and $r_\textrm{h}$=470~pc. Given the excellent observational conditions (seeing$<0.5\arcsec$), the bright RGB stars of a dwarf galaxy of this luminosity at 1.3 Mpc should have been detected. We therefore conclude that this candidate is not associated with NGC 3109 and is likely a background dwarf galaxy. While an overdensity of sources appears at $g-i\sim26$and $i\sim26$ relative to the background, the underlying unresolved light introduces high photometric uncertainties, making it difficult to determine the dwarf’s distance. Space-based imaging will be needed to determine its exact distance and properties.

\section{Completeness Tests} \label{completeness}

Understanding the completeness of the dwarf search is essential for comparing the NGC 3109 dwarf census with known or simulated satellite populations. We generate artificial dwarf galaxies and inject them into the images to measure their recovery rate. To accurately replicate all potential biases that may arise during the photometry process, we inject the dwarf galaxies into the data prior to the coaddition step or the application of the SOF photometry pipeline.

\subsection{Injecting artificial dwarfs at the image level}

\begin{figure}[t]
\centering
\includegraphics[width=0.49\textwidth]{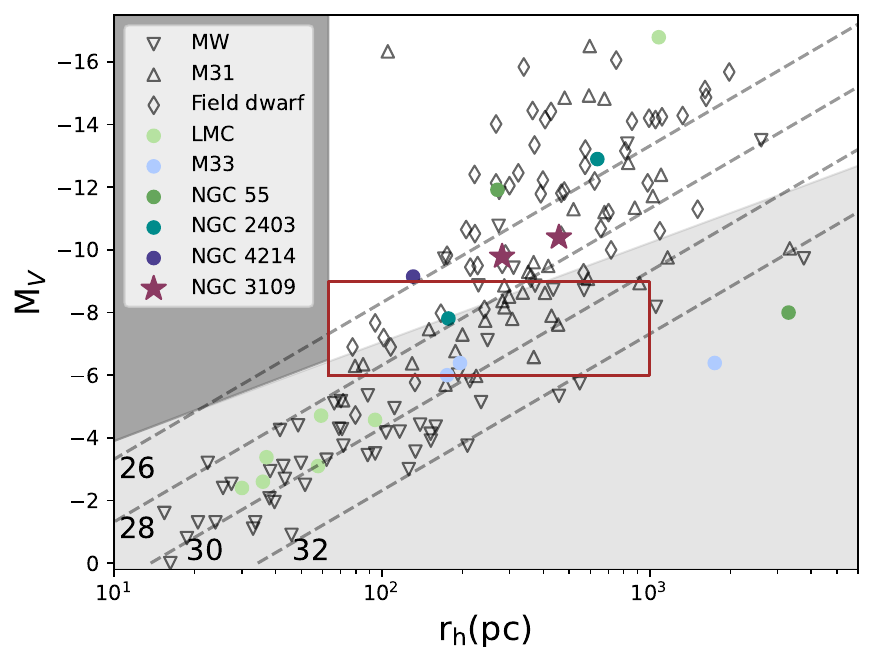}
\caption{Size-luminosity distribution of dwarf galaxies. Grey markers represent dwarfs around the Milky Way, M31, and field dwarfs from \cite{Pace2024}. %Appendix~\ref{referencemeasure} present a complete list of the references. 
We overlay the satellites of the LMC, M33, and other LMC-mass hosts as colored circles. Red stars represent known satellites of the SMC-mass host NGC 3109. Constant line of surface brightness within the half-light radius are marked by grey dashed lines. The brown selection box represents the region where detection limits are determined. The white region corresponds to the area where the dwarf sample is estimated to be $>$50\% complete. The light grey region indicates where dwarf galaxies are more difficult to detect or remain undetected. The dark grey area represents a zone occupied by compact stellar objects, where we do not expect to detect any dwarf galaxies.}
\label{size_lumi}
\end{figure}

\quad Injecting simulated dwarf galaxies into the images and generating the updated photometric catalog is computationally intensive. We therefore focus on the parameter space where known dwarf galaxies exist. The detectability of a dwarf galaxy primarily depends on its luminosity and size, which relate to its surface brightness \citep{Koposov2008, Drlica2020, Doliva2022}: we thus select the range of interest for magnitude to be $-6<M_V<-9$. The corresponding half-light radius of a faint dwarf is mainly in the range of 63$<r_\mathrm{h}$(pc)$<$1000 \citep[$1.8<\log(r_\mathrm{h}(\mathrm{pc}))<3$;][]{Brasseur2011,Doliva2023}. We assume the age, distance, and metallicity to be of secondary importance for the detection of a dwarf and we fix these parameters at 10 Gyr, 1.3 Mpc, and $-2.0$, respectively \citep{Weisz2019}. Detection limits can be extrapolated beyond the size-luminosity range explored here, as illustrated in Figure~\ref{size_lumi}.

\quad Stellar photometry in the $g$ and $r$-bands is simulated from Padova isochrones and luminosity functions \citep{Bressan2012} based on the method described in \cite{Doliva2022}. Since we inject the dwarfs directly at the image level, photometric uncertainties and stellar completeness are naturally accounted for. However, to speed up the injection process, we replace the flux contributed by stars fainter than 28th mag in the $g$-band with a diffuse light component, an exponential model of appropriate total flux and size for the dwarf.

\quad After simulating the CMDs of fake dwarf galaxies, we inject fake dwarfs into our images using a synthetic source injection (SSI) pipeline built for DELVE analyses of coadded catalogs \citep{Anbajagane2025b}, and this pipeline follows the approach of the \textit{Balrog} SSI pipeline from the Dark Energy Survey \citep{BalrogY1,Everett2022,Anbajagane2025}. In our SSI pipeline, each dwarf is modelled using an exponential profile for both its unresolved and resolved components. The resolved component is modelled as a distribution of point sources with a number density following an exponential profile. Both components follow the ellipticity assigned to the dwarf, and are randomly rotated prior to injection. The dwarfs are placed on a hexagonal spatial grid within each tile to prevent any overlap between the injected objects. The spacing of the grid is set by the half-light radius of the largest dwarf being injected in a given tile. Once a given dwarf object is assigned to be injected into a coadd tile, its magnitudes are reddened using interstellar extinction coefficients from \cite{Schlegel1998}, using $A_V$ factors from \cite{Abbott2021}. The object is then injected into every CCD image used in obtaining the coadded tile. For each CCD image, the object is convolved with the image’s PSF model and its signal is modulated by the estimated zeropoint of the image. The \textsc{Galsim} package \citep{Galsim} is employed to simulate both the diffuse and resolved components. The resulting source-injected images are then processed through the same pipeline as that used on the data (Section 3) to obtain coadd images and SOF object catalogs. By injecting into the real images, we accurately capture the noise, background, observing conditions, and artifact properties of each image. Figure~\ref{imagecandidate} presents resulting images for two of our artificial dwarfs.

\subsection{Detection limits} 

\begin{figure}[t]
\centering
\includegraphics[width=0.48\textwidth]{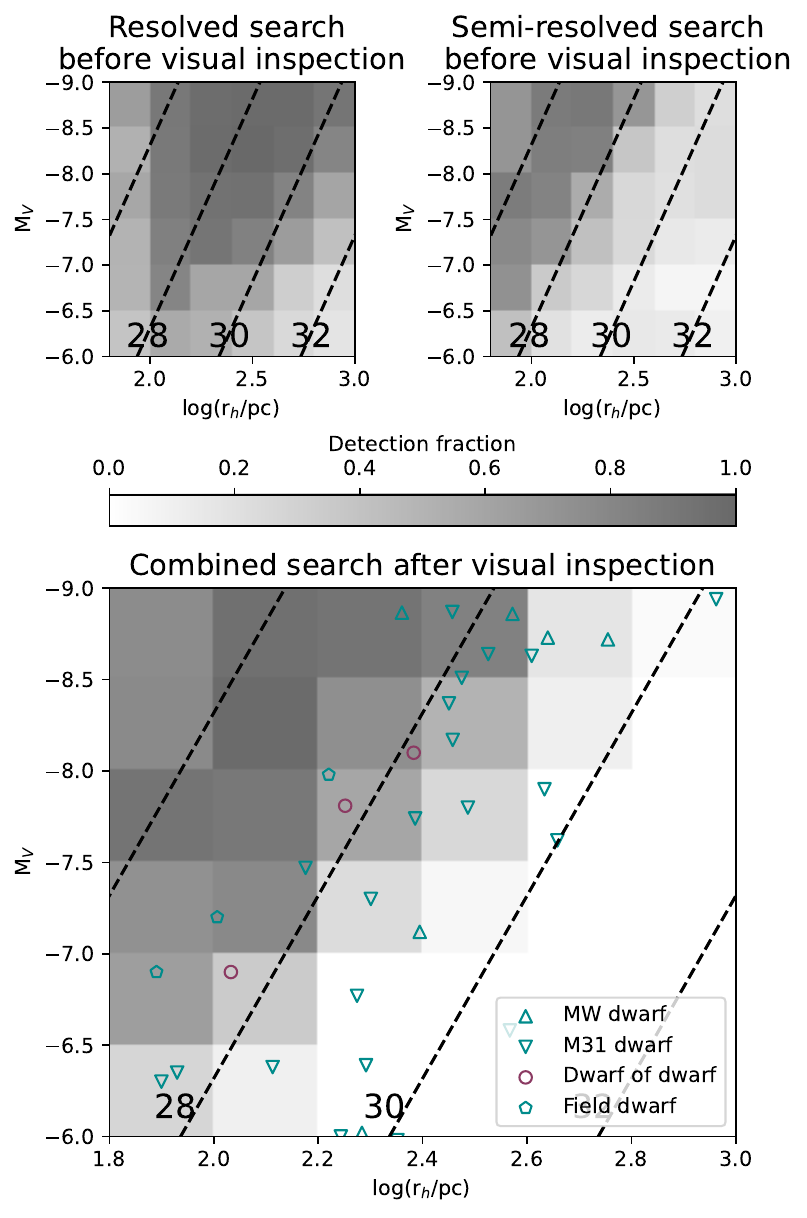}
\caption{Completeness of the automatic search for NGC 3109 dwarf galaxy satellites as a function of satellite size and magnitude. \textit{Top panels:} Detection limits for each method used in this paper—resolved and semi-resolved before visual inspection. We highlight the complementarity of those methods: the resolved method is more sensitive to extended dwarfs, while the semi-resolved method better detects compact ones.
\textit{Bottom panel:} Overall detection limits for the NGC 3109 search after visual inspection. Constant line of surface brightness within the half-light radius are marked by grey dashed lines. We are complete down to a surface brightness of $\sim 28$ mag\ arcsec$^{-2}$ and sensitive down to a surface brightness of $\sim 29.5$ mag\ arcsec$^{-2}$. Local Group dwarf galaxies are overlaid, demonstrating sensitivity to the surface brightness range relevant for detecting dwarf galaxies. Both Antlia and Antlia B lie outside the magnitude-size range explored in this completeness study, with $M_V<-9.0$.} 
\label{deteclimitstot}
\end{figure}

\quad We inject a total of $\sim$ 1,800 fake dwarf galaxies to determine the completeness of our survey. We focus on five representative fields selected to span the range of image quality (i.e., 10$\sigma$ depths) and distributed across the full extent of the survey area.

We then apply both the resolved and semi-resolved search techniques as described in Section~\ref{searchalgo} to understand our detection fraction. To estimate the real ratio of dwarf galaxies that would be detected per bin of size and magnitude, we follow the same process as for the search, and we inspect visually a representative sample through Zooniverse which is mixed with the detections from the dwarf search. The final detection limits are obtained by calculating a weighted mean of the detection limits for each field, reflecting the distribution of field completeness across the survey. 

\quad Figure~\ref{deteclimitstot} presents the completeness of the search for dwarf galaxies around NGC 3109 with the combination of the resolved and semi-resolved searches. Each size-luminosity bin contains 50 dwarf galaxies. Known dwarfs from \cite{Pace2024} are overlaid on this plot, highlighting that we are sensitive to the surface brightness of interest here. We are complete to a surface brightness within the half-light radius of $\sim 28$ mag\ arcsec$^{-2}$ and sensitive down to $\sim 29.5$ mag\ arcsec$^{-2}$. We note that the resolved and semi-resolved methods are complementary (Figure~\ref{deteclimitstot}), and employing both enables an effective search for dwarf galaxies around NGC 3109. While the resolved method appears more effective for more extended dwarfs, the semi-resolved approach allows us to identify compact dwarfs, which are more heavily affected by crowding. We encourage reader interested in using our detection limits to access \url{https://github.com/dolivadolinsky/detection_limits}. 

\section{NGC 3109 satellite system properties} \label{discussion}
\subsection{Luminosity function} \label{sectionLF}

\begin{figure}[t]
\centering
\includegraphics[width=0.49\textwidth]{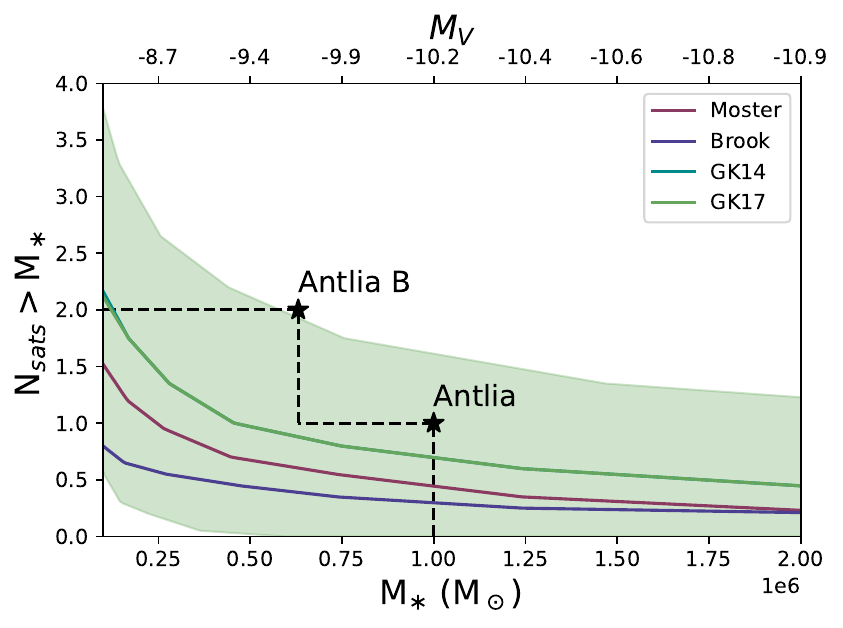}
\includegraphics[width=0.489\textwidth]{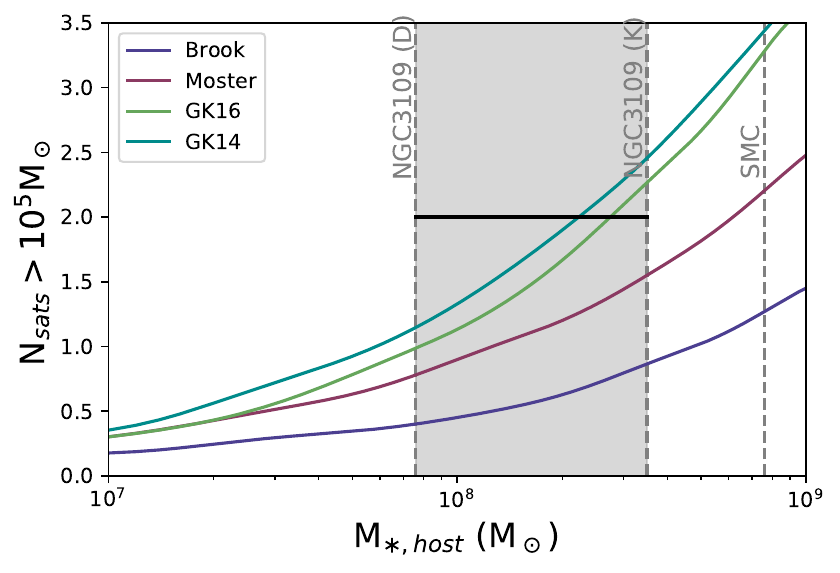}
\caption{\textit{Top panel:} Cumulative distribution of satellites as a function of their stellar mass. Theoretical predictions for a SMC-mass host are taken from \citet{Dooley2017a}. Full color lines are the mean number of satellites expected from the dark matter \textit{Caterpillar} simulation suite \citep{Griffen2016}, coupled with different stellar-mass halo ratios \citep{Moster2013,Brook2014,GK2014,GK2017}. The green shaded region represent the halo-to-halo scatter when using the GK17 relation. The observed luminosity function of NGC 3109 is superimposed in black, we did not apply any correction for incompleteness. \textit{Bottom panel:} Number of expected satellites with M~$>10^5$$M_\odot$ depending on the stellar mass of the host as extracted from \cite{Dooley2017b}. For a description of the theoretical predictions, see the caption of the upper panel. We represent here two values for NGC 3109 stellar mass, one extracted from \cite{Karachentsev2013}, marked K, and one from \cite{Dooley2017b}, marked D. The solid black line shows the number of NGC 3109 satellites with M~$>10^5$$M_\odot$.}
\label{lfunction}
\end{figure}

\quad In the top panel of Figure~\ref{lfunction}, the dashed black line shows the observed luminosity function for NGC 3109, which includes the two confirmed dwarf satellites. We overlay theoretical predictions for an SMC-mass host from \citet{Dooley2017a}, which estimate a fiducial number of satellites with $M_\star > 10^5 M_\odot$ within the host's virial radius, ranging from approximately 0.5 to 3.5. These predictions are based on the dark matter-only \textit{Caterpillar} simulations \citep{Griffen2016}, combined with four different stellar-halo mass relations (SHMRs) from \cite{Moster2013, Brook2014, GK2014, GK2017}. The completeness of our survey, as calculated previously, indicates that we are 80\% complete down to $M_V=-8.0$ (assuming a mass-to-light ratio of 1, this corresponds to $M_\star \sim 10^5 M_\odot$), enabling a nearly direct qualitative comparison between the simulated and observed luminosity functions within this mass range. The observed luminosity function tends to lie on the higher end of predictions at the bright end, with two known satellites having $M_V >-9$, compared to a fiducial prediction of $\sim$1. However, when accounting for the halo-to-halo scatter, the observations still fall within the 1-sigma range of the predictions made using the SHMR from \cite{GK2017}. It is important to note that our search covered only $\sim70$ kpc of the $90$ kpc virial radius of the halo, leaving approximately $30\%$ of the virial volume unexplored, which may result in a slight underestimation of the number of observed dwarf galaxies. However, the fraction of missed dwarfs due to spatial coverage should be lower, as they tend to be more centrally concentrated \citep{Dooley2017b}.

\quad The theoretical number of satellites shown in the top panel of Figure~\ref{lfunction} may be slightly overestimated because it is based on the more massive SMC-mass host (see Table~\ref{tab:targetlist}). The bottom panel of Figure~\ref{lfunction} shows the average expected number of satellites with M$_\star~>10^5$ M$_\odot$ as a function of host stellar mass, following \cite{Dooley2017b}. In \cite{Dooley2017b} simulations, halo masses are estimated from stellar masses assuming an SHMR which results in a wide range of values. On the observational side, determining the halo mass of NGC 3109 is challenging due to its edge-on morphology \citep{Li2020}, and there is a wide range of stellar mass estimates in the literature, as discussed in \cite{Garling2024}. We present two of those values in Figure~\ref{lfunction} to give a representative range. One is estimated from the K-band magnitude assuming a mass-to-light ratio of 1 resulting in a stellar mass of NGC 3109 of $3.5\times10^8 M_\odot$ \citep{Karachentsev2013}. The other one is obtained by fitting the stellar mass-to-light ratio to the radial velocity profile of NGC 3109 which results in a stellar mass of $7.6\times10^7 M_\odot$ \citep{Dooley2017b,McConnachie2012,Blais-Ouellette2001}. The uncertainty surrounding both the halo and stellar mass measurements of NGC 3109 complicates quantitative comparisons with theoretical models. Fortunately, upcoming MADCASH$+$DELVE-DEEP papers will address this issue by studying the host dwarf stellar halo and, in particular, globular clusters, which will provide more reliable mass estimates. Using the value from \cite{Karachentsev2013}, the predicted number of satellites matches expectations. The estimate from \cite{Dooley2017b} suggests that NGC 3109 hosts more bright satellites with M$\star > 10^5$ M$\odot$ than the average predicted by simulations. However, assuming a halo-to-halo scatter similar to the top panel of Figure~\ref{lfunction}, the observations for both mass ranges remain within the 1-sigma predictions based on the SHMR from \cite{GK2017}.

\quad Recent complementary surveys, ID-MAGE and ELVES-Dwarfs, have also investigated satellite systems around MC-mass hosts. ID-MAGE examined 36 MC-mass hosts, including 26 of SMC-mass, while ELVES-Dwarfs focused on 8 MC-mass hosts, 6 of which are SMC-mass. In both surveys, the hosts are located at greater distances ($4 < D < 12$ Mpc), and satellites were detected via integrated light, resulting in a brighter completeness limit of $M_V \sim -9$. ID-MAGE reported an average of $2 \pm 0.6$ candidate satellites per host with $M_V \lesssim -9$, with distance confirmation efforts currently underway. ELVES-Dwarfs found between 0 and 2 satellites per host above a similar brightness threshold, with most distances confirmed using surface brightness fluctuation measurements. These findings support the consistency of NGC~3109 satellite luminosity function not only with theoretical predictions (Figure~\ref{lfunction}) but also with the observed luminosity function for SMC-mass hosts from these two recent surveys.

\subsection{Stellar mass gap}
\quad The two bright satellites around NGC 3109 draws our attention to the small stellar mass gap between the host dwarf galaxy and its most massive satellite. According to $\Lambda$CDM models, the mass distribution of satellites around LMC- or SMC-sized hosts is relatively smooth, so the existence of a satellite within three magnitudes of the host is expected \citep{Deason2013,Sales2013}. \cite{Patel2018} discuss the significant stellar mass gap of over four orders of magnitude between M33 and its largest satellite, And XXII. While the LMC has at least one bright companion, the SMC \citep{Patel2020}, the next most massive satellites are ultra-faint dwarf galaxies, making the stellar mass gap significant \citep{Patel2020}. However, studying satellites of M33 and the LMC/SMC pair is challenging because they have fallen into the dark matter halos of the MW and M31. To conclusively address this potential lack of bright satellites, it is essential to increase the sample size of satellite systems around dwarf galaxies. Among our MADCASH$+$DELVE-DEEP host dwarfs, the three studied hosts all exhibit a difference of three orders of magnitude or less between the stellar mass of the host and its brightest satellite, consistent with expectations from cosmological simulations. NGC 2403 \citep{Carlin2024} has a stellar mass gap of about two orders of magnitude with its most massive satellite, DDO 44. In the case of NGC 55 \citep{Medoff2025}, ESO294-010 lies just within its virial radius with a mass within 3 orders of magnitudes of its host dwarf. Finally, NGC 3109 has a stellar mass gap of less than two orders of magnitude with both Antlia and Antlia B. While those recent observations of bright satellites are more consistent with $\Lambda$CDM predictions, a larger sample of host dwarf systems with deep surveys is needed for a more thorough comparison of observed and predicted satellite mass distributions.

\subsection{NGC 3109 environmental influence}

\quad Figure \ref{size_lumi} presents the size-luminosity distribution of dwarf galaxies in various environments, including those surrounding the MW, M31, and field dwarfs cataloged by \cite{Pace2024}. We specifically highlight satellites of LMC-mass hosts, noting that while ultra-faint dwarf galaxies are detectable around the LMC and M33, observational limitations hinder the detection of such low-luminosity systems around more distant dwarf hosts. NGC 3109, the only SMC-mass host included in this figure, has two known satellites, Antlia and Antlia B. Both Antlia and Antlia B closely follow the established size–luminosity relation observed in faint dwarf galaxies across a range of environments.  The derived luminosities and structural parameters of these satellites do not show clear indications that the gravitational influence of NGC 3109 has significantly affected their structure.

\quad Both Antlia and Antlia B have been the subject of detailed studies on their star formation histories, based on HST observations from \cite{McQuinn2010} and \cite{Hargis2019}, which show that both dwarfs have continued forming stars until very recently. However, both Antlia and Antlia B show little to no star formation in the past 100 Myr \citep{McQuinn2010,Hargis2019}. Their \hi\ gas fractions are similar to those of low-mass dwarf irregulars in the Local Volume \citep{Sand2015}, but are lower than that of the nearby isolated dwarf Leo P \citep{McQuinn2015}, and significantly lower than the typical values observed in isolated dwarfs with slightly higher stellar masses \citep[$10^7~M\odot< M\ast<10^{8.6}~M_\odot$][]{Bradford2015}. While \cite{Jahn2022} suggest that ram pressure stripping is the primary quenching mechanism for SMC-mass hosts, \cite{Garling2024} argue that recent quenching in these dwarfs is mainly due to gas depletion from star formation-driven outflows, with a secondary contribution from ram pressure stripping. We note that tidal interactions are not uncommon among dwarf galaxies around LMC-mass hosts, as discussed in \cite{Carlin2024}. In the case of SMC-mass hosts, evidence of tidal interactions between Antlia and NGC 3109 has also been reported \citep{Penny2012}, supported by the presence of an extended stellar halo feature \citep{Hidalgo2008} and a distorted \hi\  disk \citep{Carignan2013} in NGC 3109.

\section{Conclusion} 

\quad In this paper, we conduct the first systematic search for satellites around the SMC-mass galaxy NGC 3109, located at a distance of 1.3 Mpc. We employ two different search methods: a matched-filter approach to detect resolved features through overdensities of resolved stars and cuts on SExtractor parameters to identify semi-resolved/crowded ones. After visually inspecting approximately 3,000 detections, we did not identify any other high-probability satellite candidates. We found a background dwarf galaxy requiring space-based follow-up to confirm its distance. We determine the detection limits of our dwarf search as a function of luminosity and size, and find that we are complete until a surface brightness of $\sim 28$ mag\ arcsec$^{-2}$ and sensitive down to $\sim 29.5$ mag\ arcsec$^{-2}$. In other words, after convolution of the detection limits with a size-luminosity relation \citep{Brasseur2011}, we are $\sim 80\%$ complete down to $M_V\sim-$8.0 and sensitive down to $M_V\sim-$6.0. We therefore now have a well-understood sample of NGC~3109 satellite galaxies. We use these results to derive the NGC 3109 luminosity function and put its satellite system into context:
\begin{itemize}
    \item The derived luminosity function of NGC~3109 lies on the high end of prediction by cosmology and galaxy formation models. It includes two satellites with $M_V > -9$, whereas simulations for an SMC-mass host generally predict only one in this range. However, this result remains within expectations when accounting for the halo-to-halo variation.
    \item NGC 3109 has two bright satellites within three orders of magnitude of its mass, aligning with the predictions by $\Lambda$CDM. 
    \item The star formation properties of NGC 3109’s satellites suggest that an SMC-mass host exerts a weak environmental influence on its satellite galaxies. 
\end{itemize}
\quad With data on only one SMC-mass system and two well-studied objects around NGC 3109, our conclusions about the environmental influence of SMC-mass systems - as well as comparisons between this observed satellite system, other systems, and theoretical predictions—must remain qualitative for now. Small-number statistics are known to introduce significant biases, limiting quantitative comparisons. Nevertheless, these results highlight the importance of rigorously investigating the satellite systems of LMC- and SMC-mass hosts. Addressing the challenge posed by small-number statistics requires extensive efforts in identifying and characterizing satellites across multiple dwarf galaxies. This is the primary goal of the MADCASH+DELVE-DEEP surveys, with a series of papers published or in preparation to advance this work \citep{Sand2015,Carlin2016,Carlin2021,Fielder2025,Medoff2025}. With upcoming publications, a total of 11 satellite systems will be analyzed, significantly expanding the sample of known objects. Furthermore, the next generation of surveys and telescopes—such as LSST \citep{Ivezic2019}, Euclid \citep{Euclid2022}, and Roman \citep{Roman2019}—will considerably expand the search for semi-resolved dwarf galaxies in the Local Volume \citep{Mutlu-Pakdil2021,Jones2024}. These advances are expected to yield numerous dwarf-of-dwarf discoveries, which in turn will lead to an increase of the number of MC-mass systems studied with deep observations. This will enable robust quantitative comparisons between simulations and observations of MC-mass satellite systems, deepening our understanding of galaxy formation and evolution.

\section*{Acknowledgments}
\begin{acknowledgments}
We thank Kristen McQuinn for insightful discussions.

Research by ADD, DC, and SL was supported by NSF grant AST-1814208. DJS and the Arizona team acknowledges support from NSF grant AST-2205863. KS acknowledges support from the Natural Sciences and Engineering Research Council of Canada (NSERC). Research by AP was supported by NSF grand AST-2008110.

The DELVE project is partially supported by Fermilab LDRD project L2019-011 and the NASA Fermi Guest Investigator Program Cycle 9 No. 91201. The DELVE project is partially supported by the National Science Foundation under Grant No. AST-2108168, AST-2108169 and AST-2307126.

This project used data obtained with the Dark Energy Camera (DECam), which was constructed by the Dark Energy Survey (DES) collaboration. Funding for the DES Projects has been provided by the US Department of Energy, the U.S. National Science Foundation, the Ministry of Science and Education of Spain, the Science and Technology Facilities Council of the United Kingdom, the Higher Education Funding Council for England, the National Center for Supercomputing Applications at the University of Illinois at Urbana-Champaign, the Kavli Institute for Cosmological Physics at the University of Chicago, Center for Cosmology and Astro-Particle Physics at the Ohio State University, the Mitchell Institute for Fundamental Physics and Astronomy at Texas A\&M University, Financiadora de Estudos e Projetos, Fundação Carlos Chagas Filho de Amparo à Pesquisa do Estado do Rio de Janeiro, Conselho Nacional de Desenvolvimento Científico e Tecnológico and the Ministério da Ciência, Tecnologia e Inovação, the Deutsche Forschungsgemeinschaft and the Collaborating Institutions in the Dark Energy Survey. The Collaborating Institutions are Argonne National Laboratory, the University of California at Santa Cruz, the University of Cambridge, Centro de Investigaciones Enérgeticas, Medioambientales y Tecnológicas–Madrid, the University of Chicago, University College London, the DES-Brazil Consortium, the University of Edinburgh, the Eidgenössische Technische Hochschule (ETH) Zürich, Fermi National Accelerator Laboratory, the University of Illinois at Urbana-Champaign, the Institut de Ciències de l’Espai (IEEC/CSIC), the Institut de Física d’Altes Energies, Lawrence Berkeley National Laboratory, the Ludwig-Maximilians Universität München and the associated Excellence Cluster Universe, the University of Michigan, NSF NOIRLab, the University of Nottingham, the Ohio State University, the OzDES Membership Consortium, the University of Pennsylvania, the University of Portsmouth, SLAC National Accelerator Laboratory, Stanford University, the University of Sussex, and Texas A\&M University.
Based on observations at NSF Cerro Tololo Inter-American Observatory, NSF NOIRLab (NOIRLab Prop. ID 2019A-0305; PI: Alex Drlica-Wagner and NOIRLab Prop. ID 2015A-0130; PI: Denija Crnojevic ), which is managed by the Association of Universities for Research in Astronomy (AURA) under a cooperative agreement with the U.S. National Science Foundation. 

Based on observations made at the international Gemini Observatory, a program of NSF NOIRLab, is managed by the Association of Universities for Research in Astronomy (AURA) under a cooperative agreement with the U.S. National Science Foundation on behalf of the Gemini partnership: the U.S. National Science Foundation (United States), the National Research Council (Canada), Agencia Nacional de Investigación y Desarrollo (Chile), Ministerio de Ciencia, Tecnología e Innovación (Argentina), Ministério da Ciência, Tecnologia, Inovações e Comunicações (Brazil), and Korea Astronomy and Space Science Institute (Republic of Korea).

Gemini GS-2025A-Q-143 data was acquired through the Gemini Observatory Archive \citep{Hirst2017} at NSF NOIRLab and processed using DRAGONS \citep[Data Reduction for Astronomy from Gemini Observatory North and South;][]{Labrie2023}.

This manuscript has been authored by Fermi Research Alliance, LLC, under contract No. DE-AC02-07CH11359 with the US Department of Energy, Office of Science, Office of High Energy Physics. The United States Government retains and the publisher, by accepting the article for publication, acknowledges that the United States Government retains a non-exclusive, paid-up, irrevocable, worldwide license to publish or reproduce the published form of this manuscript, or allow others to do so, for United States Government purposes

This work has made use of the Local Volume Database (https://github.com/apace7/local volume database)

This research has made use of NASA's Astrophysics Data System, and \texttt{Astropy}, a community-developed core Python package for Astronomy \citep{Price-Whelan2018b}. 

This work was completed in part with resources provided by the University of Chicago’s Research Computing Center.

This publication uses data generated via the Zooni-
verse.org platform, development of which is funded by
generous support, including a Global Impact Award
from Google, and by a grant from the Alfred P. Sloan
Foundation.
\end{acknowledgments}
\bibliographystyle{aasjournal}
\bibliography{refv2}

\appendix 
\section{Hubble Space Telescope follow-up} \label{HST}

\begin{figure*}[t]
\centering
\includegraphics[width=\textwidth]{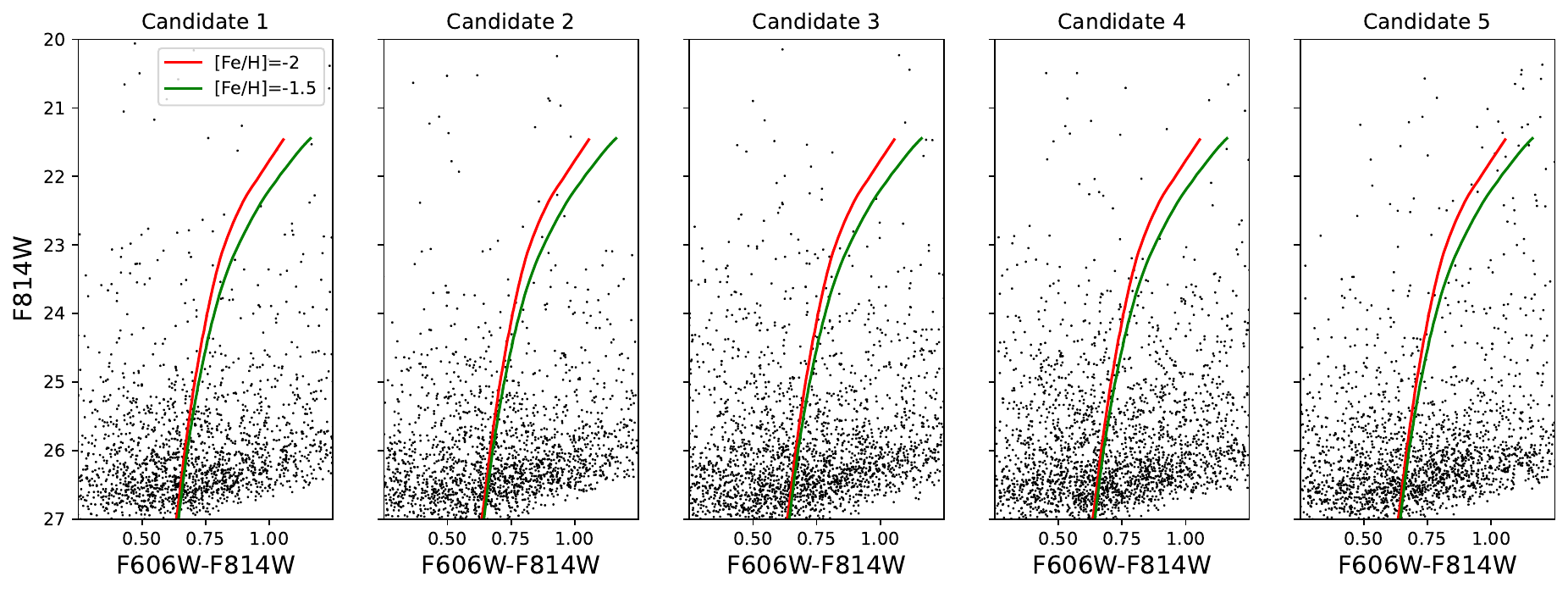}
\caption{HST CMDs of the five other candidates which were discovered via visual search in early stage of the MADCASH survey. Candidate 3 was also recovered by our semi-resolved method. We overlaid Padova isochrones for a stellar population of 10 Gyr with metallicities of $-2$ and $-1.5$, at the distance of NGC 3109. Given the absence of RGB stars at NGC 3109's distance, these detections are confirmed to be background objects.}
\label{appendixfig}
\end{figure*}

\begin{deluxetable*}{lccccc} 
\tablecolumns{3}
\tablewidth{0pt}
\tablecaption{Position of the rejected candidates followed-up with HST.\label{appenditab}}
\tablehead{\colhead{Candidate} & \colhead{R.A. (J2000)} & \colhead{Dec (J2000)}} 
\startdata
Candidate 1 & 09h57m58.7s & -22d53m29.1s \\
Candidate 2 & 09h54m51.7s & -24d22m46.0s \\
Candidate 3 & 09h51m39.6s & -24d47m5.7s  \\
Candidate 4 & 09h51m58.4s & -27d47m40.6s \\
Candidate 5 & 10h15m12.3s & -27d45m11.3s  \\
\enddata
\end{deluxetable*}

\begin{figure*}[t]
\centering
\includegraphics[width=0.26\textwidth]{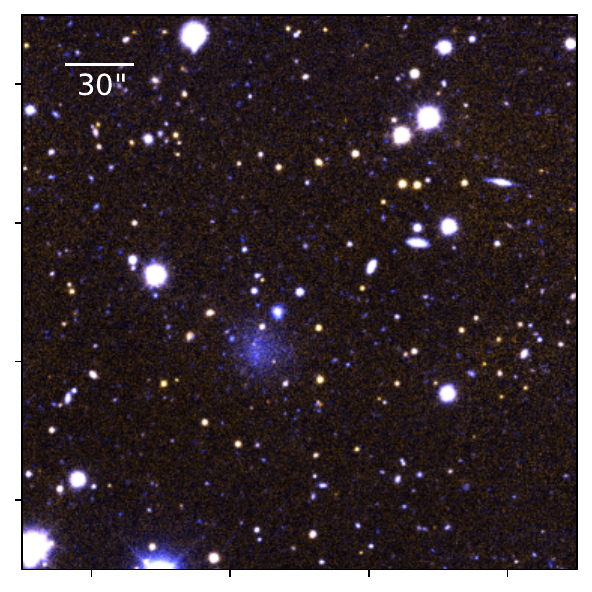}
\includegraphics[width=0.26\textwidth]{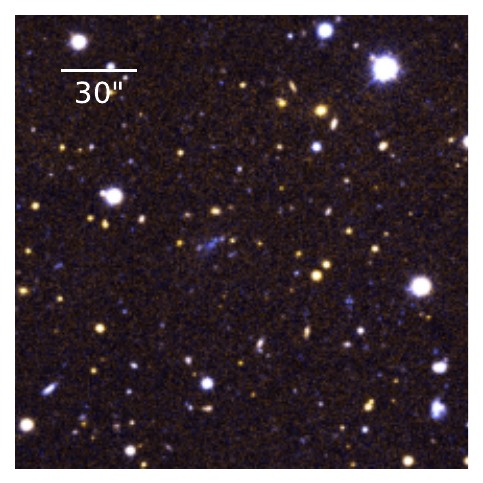}
\includegraphics[width=0.26\textwidth]{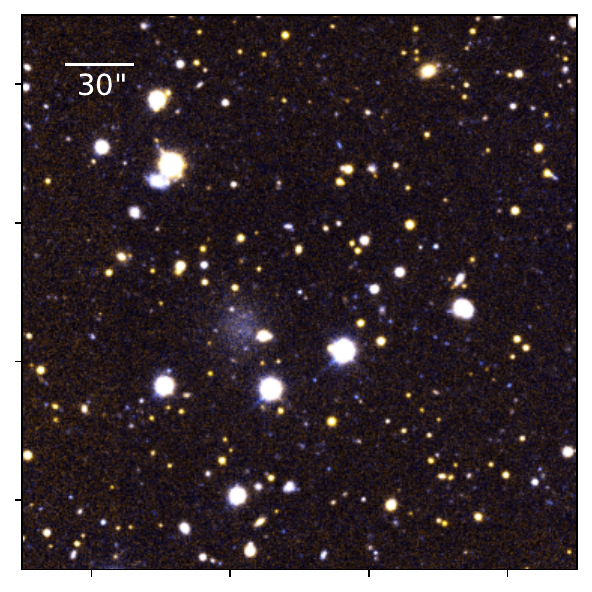}
\includegraphics[width=0.26\textwidth]{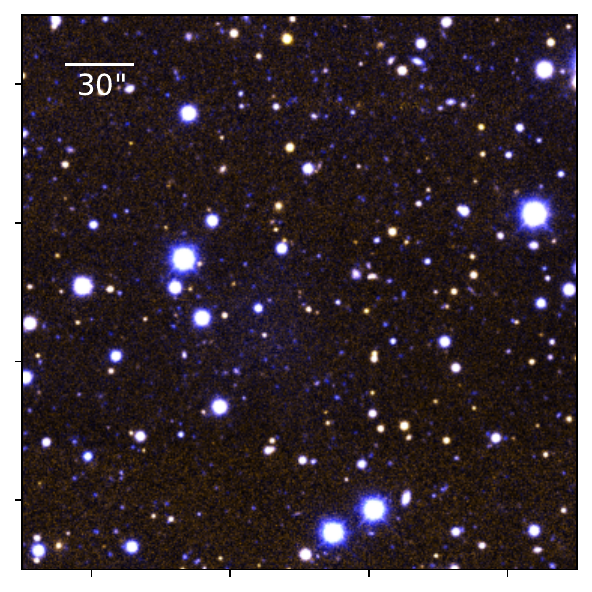}
\includegraphics[width=0.26\textwidth]{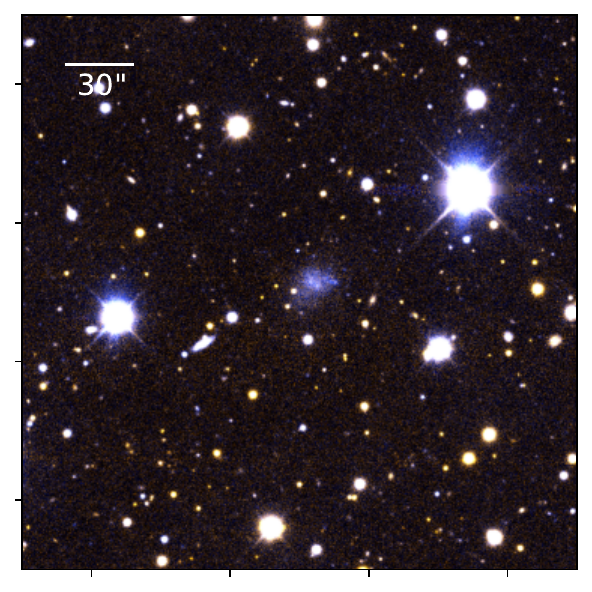}
\caption{DECam g, r, i image of five visually identified candidates (1–5, left to right, top to bottom)}.
\label{appendixfiga2}
\end{figure*}

\quad In this appendix, we present the five candidates that were identified through visual search, one of which was also recovered by our semi-resolved search (Candidate 3 in Figure~\ref{appendixfig} and Table~\ref{appenditab}). The DECam images presented in Figure~\ref{appendixfiga2} highlights that those are detected as unresolved objects. Those candidates were followed up under the HST program HST-GO-14078 (PI: J. Hargis), using the Wide Field Channel of the Advanced Camera for Surveys on 2016 October 25, 27, 28, 29, and 30. Observations for each candidate were made with F606W and F814W filters with respective exposures of 936s and 1140s. We applied the data reduction and photometry process outlined in section 5.1 of \cite{Mutlu-Pakdil2024} to obtain the CMD of the 5 candidates shown in Figure \ref{appendixfig}. We do not see any strong evidence of an RGB overdensity, although it is possible that such detections are hidden in the noise. We conclude that those detections are likely background objects.

\begin{figure*}[t]
\centering
\includegraphics[width=0.37\textwidth]
{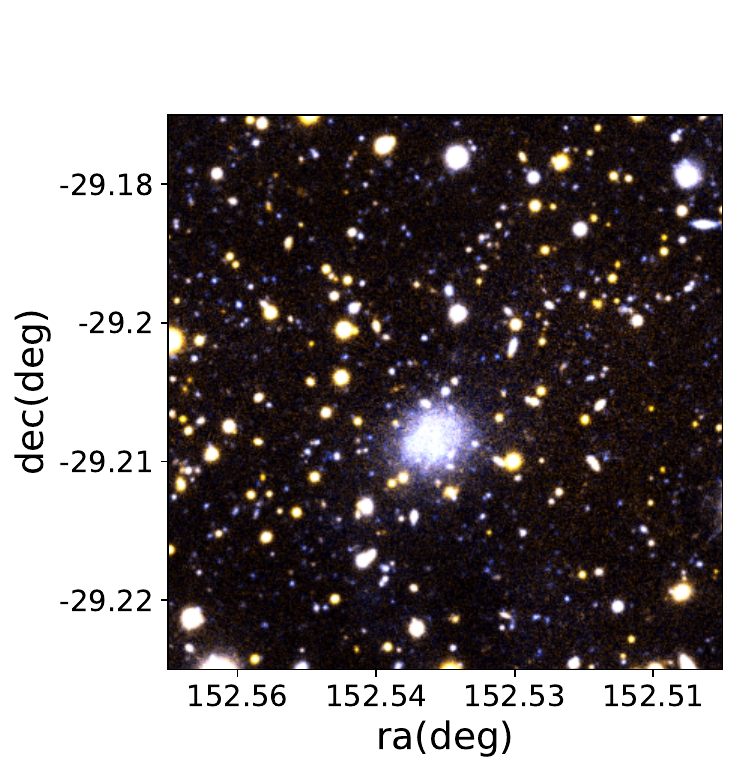}
\includegraphics[width=0.37\textwidth]{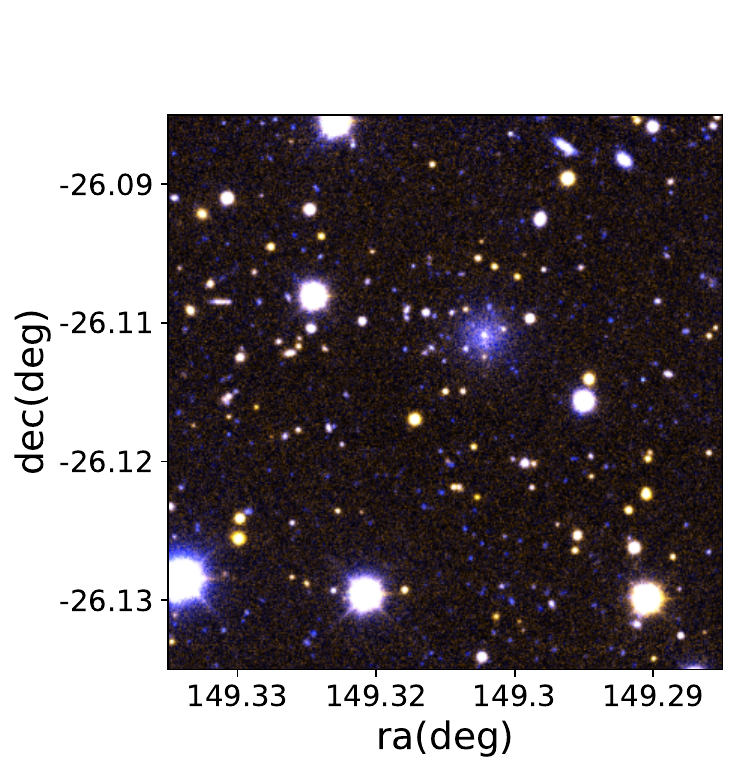}
\caption{DECam g,r,i image showing interesting objects detected during our search. We report those background detections that the algorithm identified but that did not pass the visual inspection stage, even though 2/3 of the visual inspection team rated them as candidates worthy of follow up. As they are unresolved they are most likely background dwarf galaxies.}
\label{appendixfig2}
\end{figure*}

\section{Other detections} \label{backgrounddetection}

Following our search for dwarf galaxies described in Section \ref{searchalgo}, we reviewed detections that received two out of three votes during the visual inspection step as worthy of follow-up. Here, we present two detections that stood out as possible dwarf galaxy candidates, though their fully unresolved appearance hints they are likely background objects.

%%%%%%%%%%%%%%%%%%%%%%%%%%%%%%%%%%%%%%%%%%%%%%%%

\end{document}